\documentclass[%
 reprint,
 floatfix,
superscriptaddress,
nofootinbib,
 amsmath,amssymb,
 aps,
prd,
showkeys,
]{revtex4-2}
\usepackage{booktabs}
\usepackage{centernot}
\usepackage{graphicx}
\usepackage{dcolumn}
\usepackage[dvipsnames]{xcolor}
\usepackage{bm}
\usepackage{hyperref}
\usepackage[mathlines]{lineno}
\usepackage{xcolor}
\usepackage{amsmath}
\usepackage{slashed} 

\usepackage[normalem]{ulem}

\newcommand*\mnras{MNRAS}
\newcommand*\aap{A\&A}

\newcommand*\jcap{J. Cosmology Astropart. Phys.}

\begin{document}

\preprint{APS/123-QED}

\title{Constraints on extended Bekenstein models from cosmological, astrophysical, and local data}

\author{L\'eo Vacher}
\email{leo.vacher@irap.omp.eu}
\affiliation{Institut de Recherche en Astrophysique et Planétologie, CNRS, CNES, Toulouse, France\\}
\affiliation{Universit\'e de Toulouse UPS, Toulouse, France\\}
\author{J. D. F. Dias}%
\affiliation{Centro de Astrof\'{\i}sica da Universidade do Porto, Rua das Estrelas, 4150-762 Porto, Portugal \\}
\affiliation{Faculdade de Ci\^encias, Universidade do Porto, Rua Campo Alegre, 4169-007, Porto, Portugal}
\author{Nils Sch\"oneberg}%
\email{nils.science@gmail.com}
\affiliation{Dept. F\'isica Qu\`antica i Astrof\'isica, Institut de Ci\`encies del Cosmos (ICCUB), Facultat de F\'isica, Universitat de Barcelona (IEEC-UB), Mart\'i i Franqu\'es, 1, E08028 Barcelona, Spain\\}%
\author{C. J. A. P. Martins }%
\email{Carlos.Martins@astro.up.pt}
\affiliation{Centro de Astrof\'{\i}sica da Universidade do Porto, Rua das Estrelas, 4150-762 Porto, Portugal \\}%
\affiliation{Instituto de Astrof\'isica e Ci\^encias do Espa\c co, CAUP, Universidade do Porto, Rua das Estrelas, 4150-762, Porto, Portugal}
\author{Samy Vinzl}%
\affiliation{Universit\'e de Toulouse UPS, Toulouse, France\\}
\author{Savvas Nesseris}
\affiliation{Instituto de F\'isica Te\'orica UAM-CSIC, Universidad Auton\'oma de Madrid, Cantoblanco, 28049 Madrid, Spain}
\author{Guadalupe Cañas-Herrera}
\affiliation{Leiden Observatory, Leiden University, PO Box 9506, Leiden 2300 RA, The Netherlands}
\affiliation{Lorentz Institute for Theoretical Physics, Leiden University, PO Box 9506, Leiden 2300 RA, The Netherlands}%
\affiliation{European Space Agency (ESA), European Space Research and Technology Centre (ESTEC), Keplerlaan 1, 2201 AZ Noordwijk, The Netherlands}
\author{Matteo Martinelli}
\affiliation{INAF - Osservatorio Astronomico di Roma, via Frascati 33, 00040 Monteporzio Catone (Roma), Italy}

\date{\today}

\begin{abstract}
Searching for variations of nature's fundamental constants is a crucial step in our quest to go beyond our current standard model of fundamental physics. If they exist, such variations will be very likely driven by the existence of a new fundamental field. The Bekenstein model and its extensions introduce such a scalar field in a purely phenomenological way, inducing a variation of the fine-structure constant on cosmological scales. This theoretical framework is as simple and general as possible while still preserving all the symmetries of standard quantum electrodynamics. When allowing for couplings to the other sectors of the Universe, such  as baryons, dark matter, and the cosmological constant, the Bekenstein model is expected to reproduce the low energy limits of several grand unification, quantum gravity, and higher dimensional theories. In this work, we constrain different versions of the Bekenstein model by confronting the full cosmological evolution of the field with an extensive set of astrophysical, cosmological, and local measurements. We show that couplings of the order of parts per million (ppm) are excluded for all the cases considered, imposing strong restrictions on theoretical frameworks aiming to deal with variations of the fine-structure constant.
\end{abstract}

\maketitle


\section{Introduction}
\enlargethispage*{1\baselineskip}
The fundamental constants of a given theory are quantities that can be measured but not derived within it. As such, they set the intrinsic boundaries of what a given model can explain. The contemporary standard model of particle physics contains at least 19 such parameters (a complete list can be found in \cite{UzanCst2011}). The detection of a space-time variation of one of them would be a groundbreaking discovery as it would indicate that an underlying dynamical mechanism must exist to explain their values, proving that new physics is yet to be discovered. Moreover, such a variation would be in direct violation with the universality of free fall\footnote{A variation of any of the fundamental constants would make binding energies, and thus masses of elements, space-time dependent quantities. As discussed in \cite{UzanCst2011} this implies a non-geodesic motion (regarding the Levi-Civita connexion)  depending on the composition of the element.} (in other words, the weak equivalence principle) and the local position invariance. 

According to Schiff's conjecture, this would mean a violation of the broader Einstein equivalence principle, one of the cornerstones of the general theory of relativity and, more broadly, of all metric theories of gravity (see e.g. \cite{EPs,Will2014,Will2017}).

If so, gravity could no longer be described as a geometrical phenomenon of space-time alone and/or the existence of a fifth force would be required (see e.g. \cite{fifthforcealpha}). As such, testing the stability of fundamental constants on local and cosmological scales provides a powerful test of fundamental physics beyond the reach of particle accelerators (see e.g. \citep{Martins2017review,UzanCst2011,Cosmowithvarconst2002}).

Since fundamental constants appear as theoretical foundations of a theory, the variations of the free parameters of our standard model are expected in most of the theoretical frameworks aiming to extend it, such as quantum gravity, grand unification and/or theories involving extra dimensions. One such example is the dilaton field in string theories \citep{damourEEPscalar}.

The fine-structure constant $\alpha_{\rm EM} \equiv \alpha = e^2/(4\pi \epsilon_0 \hbar c)$ is the dimensionless gauge coupling quantifying the strength of the electromagnetic interaction between charged particles. As such it can be measured very accurately using various local and astrophysical phenomena involving light.

Using a great variety of independent datasets, one can then accurately map the value of $\alpha$ across space and time (see e.g. \citep{alphacosmography}). 

While the values of fundamental constants with dimensions (e.g. $c$, $\hbar$ or $G$) are dependent of the choice of a unit system, dimensionless ratios (as gauge couplings, mass ratios, and symmetry breaking angles) will always have the same values in any units. One can indeed safely choose the natural units $\hbar=c=G=1$ while instead setting $\alpha=1$ would deeply change all the physics of the Universe.
Therefore, looking for variations of dimensionless constants is the only fully consistent approach, since their values are universal. Moreover, the dimensionless constants deeply quantify the behavior of physical phenomena.

However, from a theoretical point of view, the fine-structure constant cannot vary arbitrarily through cosmic history. Indeed, one would like to preserve fundamental symmetries of physics and their associated conservation laws such as local stress-energy conservation or gauge invariance. A safe way to do so is to implement the variation of $\alpha$ from an action principle. As originally proposed in \cite{bekensteinOriginal,Bekenstein-2}, one can promote the electric charge of the electron itself to a scalar field.
This model, called the Bekenstein model, has been further generalized accounting for interactions with matter into the Bekenstein-Sandvik-Barrow-Magueijo (BSBM) model \citep{Sandvik2002,Leal2014,Leite2016,MartinsEspresso2022} and finally by allowing for different couplings of the field with baryons, dark matter and dark energy by Olive and Pospolov (O$\&$P) in \cite{Olive2002,Alvesmartins2018}. In the later form, the model provides a very general framework to constrain variations of $\alpha$ induced by a scalar field that could be motivated by a high energy physics theory.

In the present work, we provide updated constraints on the BSBM and O$\&$P models, treating for the first time their full cosmological evolution by doing a full Bayesian Statistical analysis that combines a modified version of the \textsc{CLASS} Boltzmann-solver \cite{Lesgourgues:2011CLASS} and Monte-Carlo Markov Chains (MCMC) using \textsc{Montepython} \cite{Audren:2012wb,Brinckmann:2018cvx}.

We start by introducing the notation and theory underlying the BSBM and O$\&$P models in Sec.~\ref{sec:theory}, we then introduce the data in Sec.~\ref{sec:data-lkl}, which is later used in Sec.~\ref{sec:results} to constrain these models. Finally, we conclude by summarizing our most important results in Sec.~\ref{sec:concl}.

\section{Theoretical background} \label{sec:theory}

\subsection{Bekenstein's model and the Bekenstein-Sandvik-Barrow-Magueijo extension}

\enlargethispage*{1\baselineskip}
The original Bekenstein model introduced in \cite{bekensteinOriginal} and discussed more extensively in \cite{Bekenstein-2} seeks a purely phenomenological minimalist implementation of a varying fine-structure constant $\alpha$, that remains theoretically self-consistent with standard quantum electrodynamics (for a discussion on self-consistency of varying~$\alpha$ models see e.g. \cite{UzanCst2011} and \cite{Bekenstein-2}). 

To do so, one assumes that a variation of the electron charge\footnote{We are already here implicitly in natural units, and  considering the QED unitless gauge coupling $e/\sqrt{\hbar c \epsilon_0}$ (here in S.I. units).} $e$ is induced by a free scalar field $\epsilon$ as $e(x^\mu) \propto \epsilon(x^\mu)$. The fine-structure constant will then change according to $\alpha \propto\epsilon^{2}$. At the action level, $\epsilon$ must have a kinetic term. Its presence will also change the couplings (charges) appearing in the electromagnetic covariant derivatives, leading to a necessary redefinition of the connection coefficients $A\to \epsilon\,A$ and its associated 2-form curvature/field strength $F(A) \to F(\epsilon \,A)$. In order to preserve the gauge invariance of the theory under the unitary group $U(1)$, an extra factor of $\epsilon^{-2}\propto \alpha^{-1}$ is required in the kinetic Lagrangian density of the photon field. Such a term is formally equivalent to a space- and time-dependent change in the vacuum's permeability.

With the additional change of variable $\phi \equiv \ln(\epsilon)$, the variation of the fine-structure constant with redshift is then given by
\begin{equation}
\frac{\Delta \alpha}{\alpha_0}(z)= \frac{\alpha- \alpha_0}{\alpha_0}= \left(\frac{\epsilon}{\epsilon_0}\right)^2 - 1 = e^{2(\phi-\phi_0)}-1~,
\label{eq:alphazbsbm}
\end{equation}
with the index 0 labelling values of objects at $z=0$ and $\alpha_0 \sim 1/137$ being the value of the fine-structure constant as measured locally in the laboratory \citep{ParticleDataGroup:2020ssz}. From Eq.~\ref{eq:alphazbsbm}, one can derive the expected rate of variation of the fine-structure constant today as
\begin{equation}
\frac{1}{H_0}\left(\frac{\dot{\alpha}}{\alpha_0}\right)_{z=0}=2\phi_0'~.
\end{equation}
Hereafter primes denote derivatives with respect to $\ln(a)$ and dotted quantities refer to derivatives with respect to the cosmic time $t$.
In this basis of the field $\phi$, the full $U(1)$ invariant action for the cosmological model is given by 
\begin{align}
\label{eq:S-bsbm}
\mathcal{S}&=\int d^4x\sqrt{-g} \left[\frac{M_{*}^2}{2}\partial_\mu\phi \partial^\mu \phi
-\frac{1}{4}F_{\mu \nu}F^{\mu \nu}e^{-2\phi} \right. \nonumber \\
&-\frac{1}{2}M_{\rm Pl}^2 R  + \mathcal{L}_m + \cdots \left.\right],
\end{align}
where $R$ is the Ricci scalar, $M_{\rm Pl} = \left(8 \pi G\right)^{-1/2}$ the reduced Planck mass and we set $c=\hbar=1$. $M_*$ is a mass scale associated to the $\phi$ sector, and $F^{\mu \nu}$ is the electromagnetic field tensor associated to the connection $\epsilon A_\mu$\,.
In the present work, we will assume that $M_* = M_{\rm Pl}$\,, meaning that the energy scale of the varying constant theory is close to the one of quantum gravity as one would expect from a great unification theory.
Varying fundamental constants would also imply direct violations of the Einstein equivalence principle and/or the existence of a fifth force mediated by $\phi$ (see e.g. \cite{UzanCst2011,Will2014,Will2017}). As in \citep{Sandvik2002,Leite2016,MartinsEspresso2022} we introduce 
an additional free parameter quantifying this effect, $\zeta \equiv \mathcal{L}_{\rm EM}/\rho$, where $\rho$ is the energy density, assessing the change of electromagnetic binding energies of matter (and thus masses) in the presence of $\phi$. This $\zeta$ can be connected to the Eötvos parameter $\eta$, quantifying the violation of universality of free fall as 
\begin{equation}
\eta \sim 3\cdot 10^{-9} \zeta~.
\label{eq:eta-bsbm}
\end{equation}
As discussed in \cite{Sandvik2002}, the value and sign of $\zeta$ strongly depend on the nature of dark matter and its ability to interact with $\phi$. Extremizing the action given by Eq.~\eqref{eq:S-bsbm} with respect to $\phi$ and including this extra coupling to matter, one obtains the equation of motion for the field
\begin{equation}
\ddot{\phi} + 3H \dot{\phi} = -\frac{2}{M_{*}^2}e^{-2\phi}\zeta \rho_m~.
\label{eq:bsbm_eom}
\end{equation}
When extremizing the action with respect to the metric $g_{\mu\nu}$\,, one can derive a modified version of the Friedman equation as
\begin{equation}
H^2 = \frac{8 \pi G}{3}\left[ \rho_m(1+\zeta e^{-2\phi})+\rho_re^{-2\phi}+\rho_\phi + \rho_\Lambda \right]\,,
\end{equation}
where the field density and pressure can be deduced from the action Eq.~\eqref{eq:S-bsbm} as
\begin{equation}\label{eq:field-rho-p}\rho_\phi = \frac{M^2_*\dot{\phi}^2}{2}~, \qquad P_\phi = \frac{M^2_*\dot{\phi}^2}{2}~.
\end{equation}
From previous constraints on its coupling (e.g. \cite{MartinsEspresso2022}), we expect the contribution of the energy density of the $\phi$-field to be subdominant. As such, also its linear theory perturbations do not contribute meaningfully to the gravitational potential and can be neglected. Hence, we only show the unperturbed Friedmann equation. The same reasoning is applied to all the models considered in the present work.
\subsection{The Olive \& Pospelov extension \label{sec:O&P}}

The Bekenstein model can be generalized in a straightforward way, by letting $\phi$ be a scalar field inducing any possible variations of the fine-structure constant through a general function $\alpha \propto B_F(\phi)^{-1}$. Here again, in order to preserve gauge invariance, the field has to couple to the electromagnetic Lagrangian as
\begin{equation}
    \mathcal{L}_{EM} = -\frac{1}{4}B_F(\phi)F_{\mu\nu}F^{\mu\nu}~.
\end{equation}
A simple extension to this model is to assume that $\phi$ can have analogous couplings with all the fermion fields of the standard model $\psi$, the dark energy assumed to be a cosmological constant $\Lambda$, and a dark matter particle\footnote{The model was originally conceived with the light supersymmetric neutralino forming the WIMP.} $\chi$. We will refer to this version of the model, proposed in \cite{Olive2002}, as O$\&$P. The cosmological action becomes 
\begin{align}
    S = &\int d^4x \sqrt{-g} \ \Big[ -\frac{1}{2}M_{\rm Pl}^2 R + \frac{1}{2}M_*^2 \partial_\mu \phi \partial^\mu \phi - M_{\rm Pl}^2\Lambda_0 B_{\Lambda}\left(\phi\right) \nonumber \\ 
    &- \frac{1}{4}B_{F}\left(\phi\right)F_{\mu \nu}F^{\mu \nu}
    + \bar{\psi}\left(i\gamma^\mu D_\mu-m_\psi B_{\psi}\left(\phi\right)\right)\psi \nonumber \\
    &+ \bar{\chi}\left(i\gamma^\mu D_\mu-M_\chi B_{\chi}\left(\phi\right)\right)\chi - V\left(\phi\right) \Big]~.
\label{eq:action}
\end{align}
We will consider for this work that $V(\phi)=0$ and leave the discussion of cases with nonzero potentials for future work. As justified above, the field will be considered as homogeneous and we will not solve its perturbations equations.
The $B_i(\phi)$, $i\in [\psi,F,\chi]$ are the coupling functions of the field with the different sectors. Their deviation from $1$ encodes the strength of the scalar field coupling.
Assuming that the field value remains small on cosmological time scales, one can expand the couplings up to first order as 
\begin{equation}
\label{eq:bi}
B_{i}(\phi) = 1 + \zeta_{i}(\phi-\phi_0)~,
\end{equation}
around today's value $\Delta\phi = \phi- \phi_0 \ll 1$. This expansion is expected to be a very good approximation as the $B_i$ are already constrained to be very close to unity by observations \citep{Olive2002,Alvesmartins2018}. Given the already relatively wide allowed parameter space of the model, including the second order or higher order terms in this expansion is not necessary as their contribution to the field evolution is subdominant. Using this expansion for $B_F(\phi)$, one immediately obtains the first order evolution of the fine-structure constant with the field
\begin{equation}
    \frac{\Delta\alpha}{\alpha_0}=\frac{\alpha\left(\phi\right)}{\alpha_0}-1=B^{-1}_F\left(\phi\right)-1 = -\zeta_F \Delta\phi~,
\end{equation}
where we again Taylor expanded in $\Delta \phi$ and stopped at first order. From this expression, one can derive today's time derivative of $\alpha$ as
\begin{equation}
 \frac{1}{H_0}\left(\frac{\dot{\alpha}}{\alpha_0}\right)_{z=0} = -\zeta_F \phi_0^\prime~.
\end{equation}

As in \cite{Olive2002,Alvesmartins2018}, we will also further assume that the background cosmology evolution in the O$\&$P model remains given by the canonical Friedmann-Lemaître equation
\begin{equation}
    \left(\frac{H}{H_0}\right)^2 = \frac{8 \pi G}{3} \sum_i \rho_i\,,
\label{eq:Friedmann}
\end{equation}
where the sum extends to the field's density that remains given by Eq.~\eqref{eq:field-rho-p}. This assumption is reasonable since, as we will show, those corrections are expected to be extremely small.
Minimizing the action with respect to $\phi$ gives the coupled Klein-Gordon equation of motion
\begin{equation}
    \ddot{\phi} + 3H \dot{\phi}  = -\frac{1}{M_*^2}\sum_i \rho_i \zeta_i\,,
\label{eq:bkst_eom}
\end{equation}
where $\zeta_i = \left(\zeta_\chi,\zeta_\Lambda,\zeta_b\right)$. Note that $\zeta_F$ does not appear in equation of motion due to a null averaging of the photon fields $\langle F^2 \rangle$.

For a system of two masses of Aluminium and Platinum, the Eötvos parameter $\eta$, quantifying deviations from the weak equivalence principle can be expressed as~\cite{Olive2002}
\begin{equation}
    \eta \simeq  \zeta_p\left(\zeta_n -\zeta_p + 2.9 \cdot 10^{-2}\zeta_F\right)\,,
\end{equation}
where $\zeta_p$ and $\zeta_n$ are respectively the coupling constants of the field to protons and neutrons. To simplify the parameter space, in the following we will assume that there exists a single coupling to baryons $\zeta_b$ such that $ \zeta_p \simeq \zeta_n \simeq \zeta_b $, allowing us to write the simple expression for the Eötvos parameter $\eta$ in term of the couplings constants as
\begin{equation}
    \eta \simeq  2.9 \cdot 10^{-2} \zeta_b \zeta_F .
\label{eq:eta-O&P}
\end{equation}
Due to the degeneracies of the parameter space appearing in the observables, one can only constrain their product.
As such, we introduce the new product parameters $\eta_i$ defined as
\begin{align}
    \eta_\chi &= \zeta_F\,\zeta_\chi,\\
    \eta_b &= \zeta_F \,\zeta_b,\\
    \eta_\Lambda &= \zeta_F \,\zeta_\Lambda.
\end{align}
Since we are constraining these new product parameters instead of the $\zeta$, we will only be able to recover properly the product quantities $\zeta_F\phi_0$ and  $\zeta_F\phi^{\prime}_0$ instead of the raw field parameters themselves.

\section{Datasets and Likelihoods \label{sec:data-lkl}}

We exploit the synergy of multiple datasets and their corresponding likelihoods in order to constrain the models. All these measurements are independent and probe fundamental physics at a great variety of space-time scales. The Cosmology data sets are already implemented in the {\sc Montepython} code, while the fine-structure constant and Einstein equivalence principle likelihoods are implemented as gaussian priors.

\subsection{Cosmological datasets}

In order to constrain the background cosmology, we use the likelihood based on the Pantheon Type Ia Supernovae sample \citep{SNPantheon}. We also include large scale structures and baryon acoustic oscillation data from the {\sc BOSS} DR-12 galaxy survey \citep{DR12} as well as cosmic clocks measurements from \cite{CC2016}. All of these give sharp constraints on the possible evolution of the Hubble parameter $H(z)$.

We also include the cosmic microwave background (CMB) intensity, polarization and lensing power spectra likelihoods from the latest {\it Planck} 2018 data release \citep{Planck2018,Planck_lkl}\footnote{Likelihoods can be found on the \href{http://pla.esac.esa.int/pla/}{{\it Planck} legacy archive}.}. This likelihood is giving a unique lever arm at $z\sim 1100$, further constraining the cosmology and the scalar field evolution at very high redshift.

\subsection{Fine-structure constant and Einstein equivalence principle}

Using high-resolution spectroscopy, one can obtain very accurate measurements of $\alpha$ from astrophysical sources. Doing so is possible from the position of absorption lines of the gas along the line of sight of quasi-stellar objects (QSO or quasars) at high redshifts. The positions of the lines are expected to change with $\alpha$ in a transition-specific fashion (quantified by a so-called sensitivity coefficient) that can be disentangled from the linear effect of redshift. We use a collection of measurements of the fine-structure constant from
\citep{alphaSubaru} and \citep{alphaWebb} as well as a  recent precise and accurate measurement from the ESPRESSO spectrograph \citep{alphaespresso}.

The value of $\alpha$ at $z=0.14$ can also be inferred from abundances in the Oklo natural reactor on Earth \citep{Oklo}.
\begin{equation}
    \frac{\Delta \alpha}{\alpha_0}(z=0.14) = (0.005  \pm 0.061)\,\mathrm{ppm}\,.
\end{equation}
Laboratory atomic clock experiments can use optics to constrain the current rate of change of $\alpha$ \citep{atomicclock}, which can be expressed in a dimensionless form as
\begin{equation}
    \frac{1}{H_0}\left(\frac{\dot{\alpha}}{\alpha_0}\right)_{z=0} = (0.014 \pm 0.015) \,\mathrm{ppm}\,.
\label{eq:atomicclock}
\end{equation}
Finally, sharp constraints can be added to the models considering limits on the violation of the weak equivalence principle by the MICROSCOPE satellite testing the universality of free fall with two test bodies orbiting earth
\citep{Microscope_new}
\begin{equation}
    \eta = (-1.5 \pm 2.7)\cdot 10^{-9} \,\mathrm{ppm}\,.
\label{eq:microeta}
\end{equation}
\section{Results}\label{sec:results}
We constrain the models by sampling over their parameters using MCMC chains with {\sc montepython} \citep{Audren:2012wb,Brinckmann:2018cvx} combined with a modified {\sc Class} version \citep{Lesgourgues:2011CLASS}. A discussion of the impact of a varying $\alpha$ on cosmology can be found in \citep{Planckalphame2016,HartChluba}.
The contour plots are made using the {\sc Getdist} python package \citep{getdist}. Computations are made on the cluster of the Marseille dark energy center (mardec).

The cosmological parameters we are sampling over are the reduced baryon and cold dark matter densities $\omega_{\rm b}=\Omega_{\rm b}h^2$ and $\omega_{\rm cdm}=\Omega_{\rm cdm}h^2$, the reionization redshift $z_{\rm reio}$\,, the Hubble constant $H_0$\,, the amplitude and tilt of the primordial power spectrum $n_{\rm s}$ and $\ln(10^{10}A_{\rm s})$,
and the couplings $\zeta$ or $\eta_i$ of the Bekenstein models. We adopt flat and unbounded priors for all of these parameters.  We are additionally sampling over the 21 nuisance parameters of the {\it Planck} likelihood and the absolute magnitude $M$ of the reduced Pantheon likelihood. The $|R-1|$ convergence values, further chain information, and full corner plots can be found in appendix~\ref{sec:appendix}.

We fix the values of the initial field value and speed to zero when $z\to \infty$, since one can show that these parameter choices in the radiation era do not impact the late time evolution of the field, due to the existence of attractor behaviors. The actual value of the field $\phi_0$ (or $\zeta_F\phi_0$) and its speed $\phi^{\prime}_0$ (or $\zeta_F\phi^{\prime}_0$) are derived but not sampled over. 

\subsection{BSBM model}
\begin{figure}[t!]
    \includegraphics[width=\columnwidth]{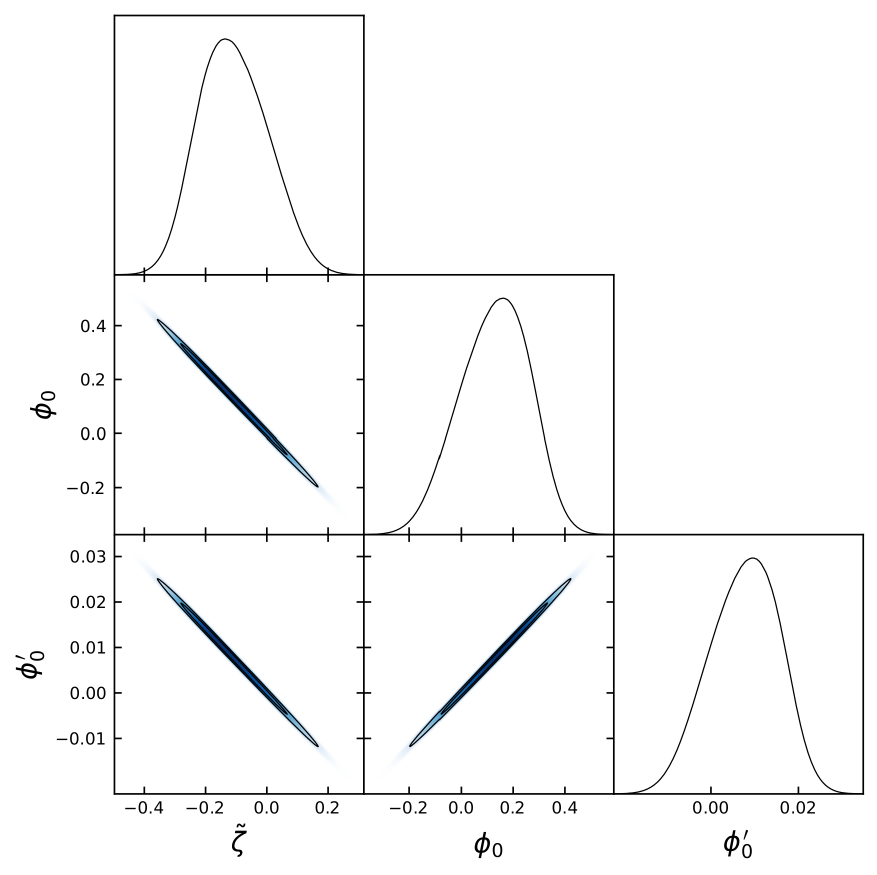}
    \caption{Contour plots for the single rescaled BSBM parameter $\tilde{\zeta}$ and the two derived parameters $\phi_0$ and $\phi^{\prime}_0$, expressed in ppm. The contours lines are representing the 68 and 95 \% confidence levels.}
    \label{fig:bsbm_short}
\end{figure}
Originally, the BSBM model has been introduced using a length scale $\omega$ to define the field units in the action (Eq.~\ref{eq:S-bsbm}) instead of the mass scale $M_{*}$ \citep{Sandvik2002}. The parameter $\omega$ is then assumed to be close to the Planck length $\omega \sim G$ \cite{Leite2016,MartinsEspresso2022}.  In our notation this would correspond to $M_{*}=1$, but we choose to instead absorb this different choice in a redefinition of the coupling constant, with $\tilde{\zeta} = 8\pi \zeta$ in order to allow for a direct comparison with previous literature.
In Fig.~\ref{fig:bsbm_short}, the derived contours of $\tilde{\zeta}$, $\phi_0$ and $\phi^{\prime}_0$ are displayed using all the likelihoods introduced in Sec.~\ref{sec:data-lkl}. 
The corresponding best-fit values and their $\sigma$ values can be found in Tab.~\ref{tab:bsbm-tab}.
We derive 
\begin{equation}
\tilde{\zeta} = -0.10^{+0.11}_{-0.13}\, \mathrm{ppm}\,.
\end{equation}
\enlargethispage*{1\baselineskip}
This result coincides with the one obtained in \citep{MartinsEspresso2022}, providing a validation of our methodology. Note that adding the recent update of the MICROSCOPE bound in the present work does not change this result. Indeed, a back to the envelope calculation combining \eqref{eq:microeta} and \eqref{eq:eta-bsbm} allows us to evaluate the width of the Gaussian prior expected from the MICROSCOPE likelihood on $\tilde{\zeta}$ to be $\sim 22$\,ppm, which is one order of magnitude larger than the one we obtained. We can hence conclude that atomic clocks measurements provide most of the constraining power on the BSBM model. For the first time however, the full model has been constrained together with the cosmological parameters and full evolution of the field right after inflation (the full plot can be found in appendix \ref{ssec:fullcorner}, in Fig.~\ref{fig:bsbm-full}). 

The only parameter $\tilde{\zeta}$ appears however to be largely uncorrelated with cosmological parameters, explaining why the two analyses lead to identical results. The field speed is constrained at one sigma as $\phi^\prime_0= (6.6^{+9.3}_{-7.3})\cdot 10^{-3}~\mathrm{ppm}$ while the field itself is constrained as $\phi_0= 0.11 ^{+ 0.16}_{-0.12}~\mathrm{ppm}$. As expected, $\tilde{\zeta}$ and the field parameters are highly correlated since they are directly related through the equation of motion (Eq.~\ref{eq:bsbm_eom}).

\begin{table}
    \centering
    \caption{Best-fit values of the BSBM parameters with associated 68$\%$ confidence levels (C.L.) in ppm.}
    \begin{tabular}{l|c}
\toprule
            \textbf{Parameter}  & 
            \textbf{68$\%$ C.L.}\\
\toprule
                {\boldmath$\tilde{\zeta}$} & $-0.093^{+0.10}_{-0.13}$ \\
\midrule
                    {\boldmath$\phi_{0}$} & $0.11^{+0.16}_{-0.12}$ \\
\midrule
                  {\boldmath$\phi^{'}_{0}$} & $0.0066^{+0.0093}_{-0.0073}$ \\
\bottomrule
\end{tabular}
\label{tab:bsbm-tab}
\end{table}

\subsection{O$\&$P model: Universal coupling to gravity}
\begin{figure*}[t!]
\includegraphics[scale=0.78]{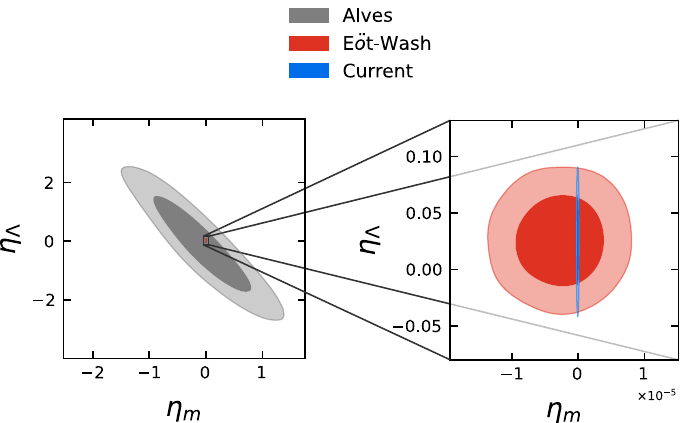}
\includegraphics[scale=0.44]{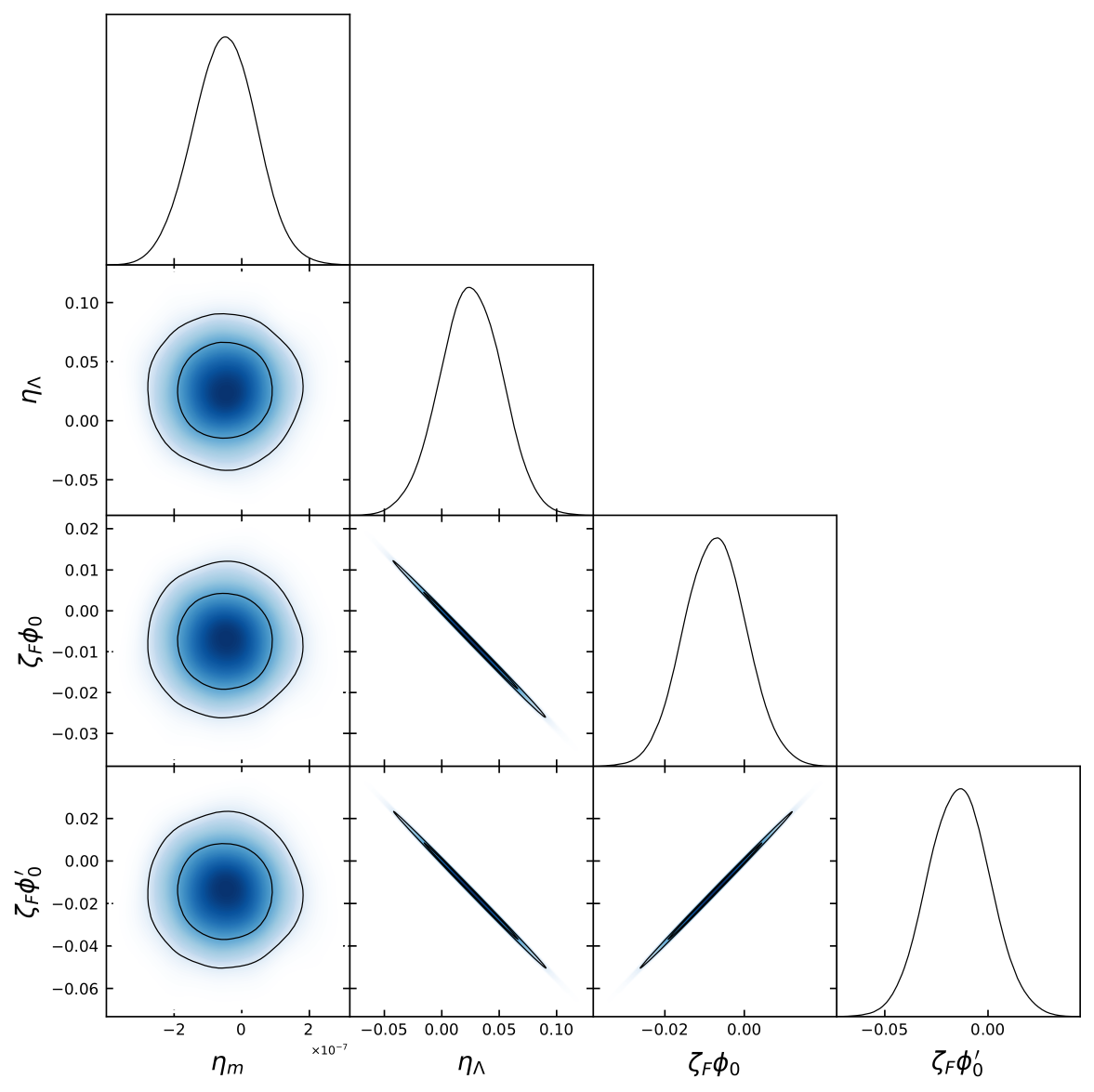}
    \caption{Left: Contour plots for the O$\&$P model under the universal matter coupling assumption with three different likelihood sets: 'Alves' (Gray), 'Eöt-Wash' (Red) and `Current' (Blue). Right: Close-up view on the O$\&$P parameter space using the `Current' likelihood set. For both plots all parameters are expressed in ppm.}
    \label{fig:zm-short}
\end{figure*}
%
Before constraining the full parameter space of the O$\&$P model, we first assume that the field couples identically to baryons and dark matter through a single parameter $\zeta_m \equiv \zeta_b=\zeta_\chi$. As in Sec.~\ref{sec:O&P}, one can then introduce the corresponding product parameter $\eta_m \equiv \zeta_F \zeta_m$. This reduction of the parameter space allows a direct comparison with previous works, such  as \citep{Alvesmartins2018}. We ran a first set of chains with all the likelihoods introduced in Sec.~\ref{sec:data-lkl} (hereafter noted 'Current') and a second one removing the MICROSCOPE prior and replacing our atomic-clock likelihood by the one used in \cite{Alvesmartins2018} and originally obtained in \citep{Rosenband} (hereafter noted `Alves'). We also consider a third situation replacing the MICROSCOPE likelihood by the  earlier measurement of $\eta$ from torsion balance by the Eöt-Wash group \citep{Eot-wash} (hereafter noted `Eöt-Wash'). This last test allows us to assess the impact of WEP tests on the parameter space and quantify the improvement brought by the recent MICROSCOPE results.

A contour plot comparison of the O$\&$P parameters in the three scenarios can be found in Fig.\ref{fig:zm-short} and the corresponding best-fits and confidence interval values are displayed in Tab.~\ref{tab:zm-tab}. As expected, the `Alves' case gives results comparable with the ones of \cite{Alvesmartins2018}, constraining the two parameters $\eta_m$ and $\eta_\Lambda$ at the ppm level, displaying a strong degeneracy between the two parameters, as they both appear on the same footing in the equation of motion (Eq.~\ref{eq:bkst_eom}). Adding a prior coming from experiments searching for violations of the WEP allows us to strongly break degeneracies as it directly constrains the coupling to matter $\eta_m$. Indeed, as shown in Fig.~\ref{fig:zm-short}, adding either the MICROSCOPE or Eöt-Wash likelihood severely restricts the otherwise very degenerate combination of $\eta_m$ and $\eta_\Lambda$. MICROSCOPE provides however sharper constraints on the matter coupling by two orders of magnitudes.
By setting the constraint 
\begin{equation}
    \eta_m = \left(-0.54^{+0.86}_{-0.90}\right) \cdot 10^{-7} \, \mathrm{ppm} \, ,
\end{equation}
this parameter relaxes its correlation with $\zeta_F\phi_0$ and $\zeta_F\phi^{\prime}_0$ and cannot significantly impact the field equation of motion anymore. 

Comparing the `Current' and `Eöt-Wash' cases clearly shows that an improvement of the accuracy of WEP measurements does not further sharpen the posterior distribution of the coupling to dark energy, which is mainly set by the atomic-clock likelihood, constraining $\eta_{\Lambda}$ at one sigma to
\begin{equation}
\eta_{\Lambda}=(0.025\pm0.027)\, \textrm{ppm}\,.
\end{equation}
Here again, as shown in Fig.~\ref{fig:zm-full} of the appendix, the constraints on the parameters of the Bekenstein field are strong enough to largely break all possible degeneracies with cosmological parameters, leaving both mostly independently constrained.
Overall, this leads to an improvement of the previous constraints of a factor of $\sim 10^8$ for~$\eta_m$ and $\sim 100$ for~$\eta_\Lambda$, considering this time the full cosmological evolution of the field with minimal assumptions. Couplings of order ppm are now excluded for this model.
\begin{table*}
    \centering
    \caption{Best-fit values of the parameters for the O$\&$P model universally coupled to matter with associated 68$\%$ confidence levels (C.L.) in ppm, from the combination of currently available data. For comparison, the analogous constraints for two earlier sets 'Alves' and 'Eöt-Wash' (see the main text) are also shown.}
    \begin{tabular}{l|c|c|cc}
\toprule
            \textbf{Parameter }  & 
            \textbf{68$\%$ C.L.} Current & \textbf{68$\%$ C.L.} Eöt-Wash  &   \textbf{68$\%$ C.L.} Alves \\
\toprule
             {\boldmath$\eta_{m}$}& $\left(\,-0.54^{+0.86}_{-0.90}\,\right)\cdot 10^{-7}$& $\left(\,-0.25^{+0.45}_{-0.44}\,\right)\cdot 10^{-5}$ &$0.05^{+0.60}_{-0.67}$ \\
\midrule
                {\boldmath$\eta_{\Lambda} $} & $0.025\pm 0.027$& $0.024^{+0.03}_{-0.027}$ & $-0.4\pm 1.1$\\
\midrule
                    {\boldmath$\zeta_F\phi_{0}$}& $-0.0073^{+0.0081}_{-0.0076}$ & $-0.007^{+0.0077}_{-0.0086}$ &$-0.7^{+9.6}_{-8.5}$ \\
\midrule
                   {\boldmath$\zeta_F\phi^{'}_{0}$}& $-0.0014^{+0.0016}_{-0.0015}$ & $-0.015^{+0.015}_{-0.017}$ & $0.17\pm 0.27$\\
\bottomrule
\end{tabular}
\label{tab:zm-tab}
\end{table*}

\subsection{Full O$\&$P model}
\begin{figure*}[t!]
\includegraphics[scale=0.5]{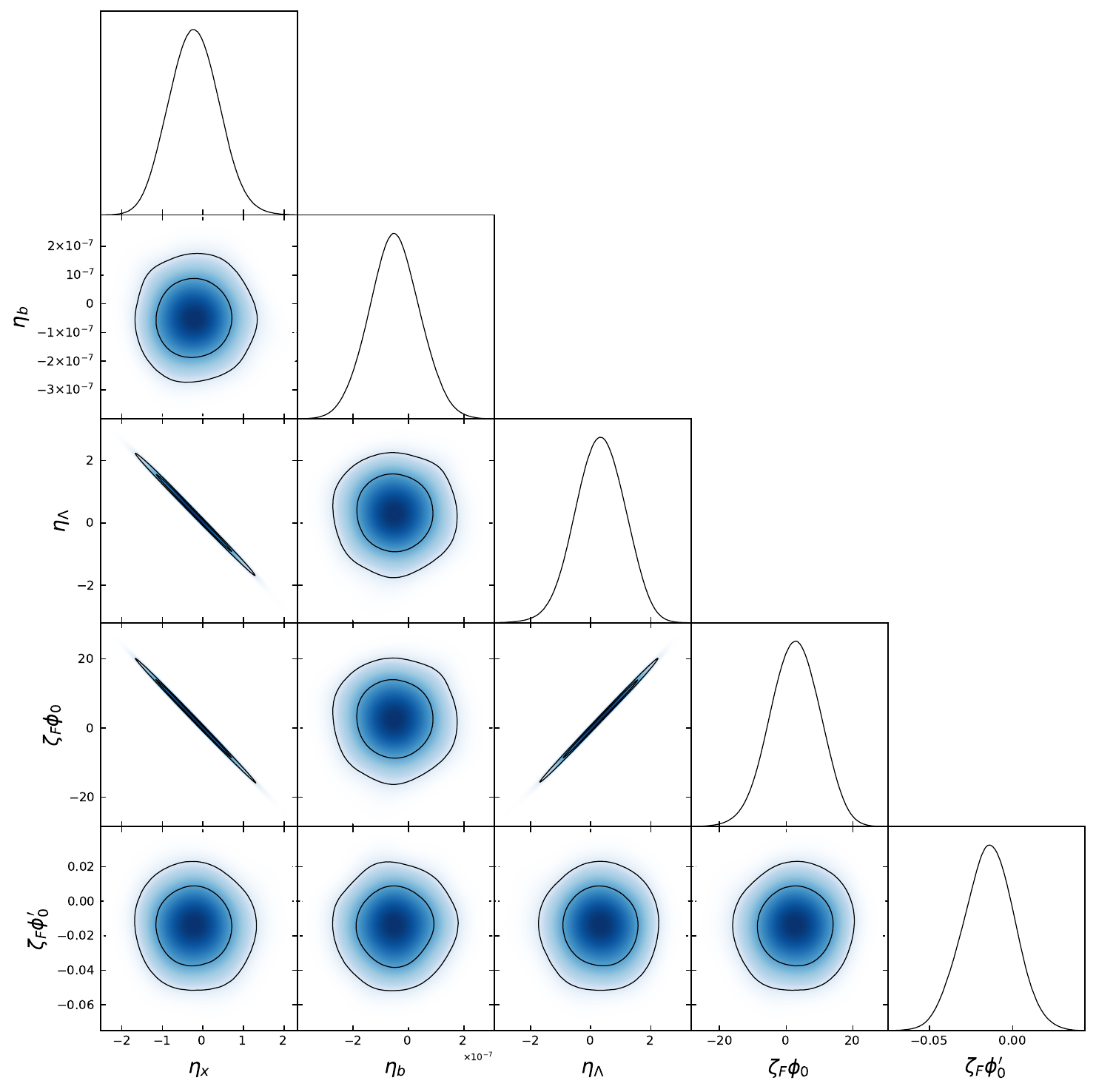}
\caption{Contour plots of the Bekenstein parameters in the full O$\&$P model in ppm.}
\label{fig:bkstfull-short}
\end{figure*}

\enlargethispage*{1\baselineskip}
We now turn to the full O$\&$P model. We propose here for the first time a constraint of its full parameter space as no previous such studies can be found in the literature. The contour plots for the field parameters can be found in Fig.~\ref{fig:bkstfull-short} and the associated best-fits with confidence levels are displayed in Tab.~\ref{tab:bkstfull-tab}. The MICROSCOPE prior is acting here directly on $\eta_b$\,, analogously to what occurred for $\eta_m$ in the previous subsection, leaving the strong degeneracy between $\eta_\chi$ and $\eta_\Lambda$. The atomic-clock likelihood indeed, is sharp enough to break the degeneracies between $\zeta_F\phi'$ and $(\eta_\chi\,,\eta_\Lambda)$, by constraining the field speed to be so small that the impact of both couplings on the speed is indistinguishable. Nevertheless, here again we see that order ppm parameters are excluded by our combination of datasets. 

As shown in Fig.~\ref{fig:bkst-full}, even this full model with a wide parameter space shows few degeneracies with the cosmological parameters, and as such it is constrained to have a minimal impact on the standard expansion history of the universe. 

As shown in Tab.~1 of \citep{Olive2002}, several models beyond the standard model of cosmology and particle physics such as Brans-Dicke, Supersymmetry, or String Theory inspired models are supposed to be contained within the framework of this extended parameter space. By excluding couplings greater than fractions of ppm, our constraints exclude their naturally expected values for most of these models.
\begin{table}
    \centering
    \caption{Best-fit values of the parameters for the full O$\&$P model with associated 68$\%$ confidence levels (C.L.) in ppm.
    }
    \begin{tabular}{l|cc}
\toprule
            \textbf{Param}  & 
            \textbf{68$\%$ C.L.}\\
\toprule
             {\boldmath$\eta_\chi$} & $-0.24^{+0.63}_{-0.66}$\\
\midrule
                {\boldmath$\eta_b$} & $\left(\,-0.54^{+0.93}_{-0.94}\,\right)\cdot 10^{-7}$\\
\midrule
              {\boldmath$\eta_\Lambda$} & $0.34^{+0.88}_{-0.85}$\\
\midrule
                    {\boldmath$\zeta_F\phi_{0}$} & $2.87^{+ 8.0}_{-7.7}$\\
\midrule
                   {\boldmath$\zeta_F\phi^{'}_{0}$} & $0.015^{+ 0.016}_{-0.015}$\\
\bottomrule
\end{tabular}
\label{tab:bkstfull-tab}
\end{table}

\section{Discussion and conclusion}\label{sec:concl}

Bekenstein models offer a very general and consistent framework for tests for variations of the fine-structure constant on cosmic scales. Even though it is expected to describe the low energy limit of several models beyond the standard models of cosmology and particle physics, it is tightly constrained by contemporary data, and expected to behave very close to the standard model. Specifically, we found that in all the generalizations considered, couplings of order of parts per million are excluded. The synergy of local, astrophysical and cosmological measurements of high precision applies an increasingly strong pressure on the credible models encompassing variations of the fundamental constants.

In this work, we constrained three implementations of the Bekenstein model. First, for the BSBM model, we have recovered, with a wider parameter space and a different methodology, the constrains derived in  \citep{MartinsEspresso2022}. This also provides a validation of our numerical analysis pipeline. Then, we constrained a variation of O$\&$P having a common coupling for baryonic and dark matter as in \cite{Alvesmartins2018}. We improve previous constraints on this model by a factor of $\sim 100$ for $\eta_\Lambda$ and $\sim$  $10^8$ for $\eta_m$. This large improvement is mainly due to the addition of very accurate local data as the MICROSCOPE prior on the universality of free fall and an improvement of the atomic-clock constrain on the time variation of the fine structure constant by an order of magnitude.

Finally, we provided a constraint of the full O$\&$P model for the first time, allowing us to exclude natural values for the couplings for almost all the high energy physics theories encompassing a varying $\alpha$ proposed in the original paper of \cite{Olive2002}, excluding a large part of the parameter space of the models. 

In all these analyses, we saw that parameters are too sharply constrained by fine structure constant and Einstein Equivalence Principle measurements to have a significant impact on cosmological evolution, as the parameters of the Bekenstein field become mostly decorrelated from cosmological ones. As it quantifies the interaction between photons and matter, it is however known that a varying $\alpha$ could have a strong impact on the physics of recombination, changing its overall duration and the width of the last scattering surface. As such, some models inspired by the Bekenstein one could significantly impact the recovered value of $H_0$ and provide ways to relax the Hubble tension (see \citep{HartChluba,Hart:2021kad}, or \cite{H0olympics} for a similar idea with the electron mass). One way to do so could be to introduce a more complex parameter space, with a different $\phi$ dependence of the couplings, the possibility of a decay of $\phi$ at intermediate times, or a non-zero potential $V(\phi)$. We note that the latter option might be subject to fine-tuning issues. As such, this kind of investigation is left for future works.

In this quest for high precision tests of fundamental physics, further progress is to be foreseen. In the long term, new experiments, under construction or being planned, will enable direct tests of the stability of fundamental constants with an accuracy never reached before. In particular the high-resolution spectrograph for the Extremely Large Telescope, formerly called ELT-HIRES and now known as ANDES \cite{HIRES} (whose Phase B of construction is starting, and expected to be operational in about 8 years) should improve the sensitivity of astrophysical measurements of $\alpha$ by at least one order of magnitude, while also extending the range of redshifts that ESPRESSO can probe. Moreover, recent theoretical and experimental developments open the possibility of improving the sensitivity of local atomic clock tests on the current drift rate of $\alpha$ by several orders of magnitude, by relying on Thorium-229 based nuclear clocks \cite{THorium}.

Constrains on the stability of fundamental constants on very large scales are also expected from wide cosmological surveys. Synergies between ground and space observations are expected from galaxy surveys performed from space with {\it Euclid} and from the ground with {\it DESI} \cite{Euclid2021,Desi2016}. Similarly, recent or incoming observations of the CMB from the ground  with telescopes such as {\it ACT} \citep{ACT}, {\it SPT} \citep{SPT} and the {\it Simons Observatory} \citep{SimonsObservatory} could surpass the last bounds set by {\it Planck} in \cite{Planckalphame2016} on the value of the fine structure constant at $z\sim 1100$. Further high precisions improvements from the CMB polarization are also to be expected in the next decades from ground with the {\it CMB Stage-4} telescope and from space with the {\it LiteBIRD} satellite \citep{CMBS4,PtepLB}.
 
The pipeline developed in the present work can easily be generalized to constrain all possible variations of the fine-structure constant driven by a scalar field and could be extended to other fundamental constants. Instead of constraining the field parameters alone, it allows to evaluate its impact in relation with all the cosmological parameters.

\section{Acknowledgments}

L.V. would like to thank J. Aumont and L. Montier for their support through all this project and A. Blanchard, B. Lamine as well as J. Lesgourgues for crucial discussions at the origin of this work.
Computations were made on the Mardec cluster supported by the OCEVU Labex (ANR-11-LABX-0060) and the Excellence Initiative of Aix-Marseille University - A*MIDEX, part of the
French “Investissements d’Avenir” program. LV would also like to thanks B. Carreres for several helps with the use of Mardec.
J.D.F.D. is supported by an FCT fellowship, grant number SFRH/BD/150990/2021. CJM acknowledges FCT and POCH/FSE (EC) support through Investigador FCT Contract 2021.01214.CEECIND/CP1658/CT0001.
NS acknowledges the support of the following Maria de Maetzu fellowship grant: Esta publicaci\'on es parte de la ayuda CEX2019-000918-M, financiada por MCIN/AEI/10.13039/501100011033. 
SN acknowledges support from the research project PGC2018-094773-B-C32, and the Spanish Research Agency (Agencia Estatal de Investigaci\'on) through the Grant IFT Centro de Excelencia Severo Ochoa No CEX2020-001007-S, funded by MCIN/AEI/10.13039/501100011033. GCH acknowledges support from the Delta Institute for Theoretical Physics (D-ITP consortium), a program by the NWO. MM acknowledges funding by the Agenzia Spaziale Italiana (ASI) under agreement n. 2018-23-HH.0.

\bibliography{apssamp}

\begin{thebibliography}{48}%
\makeatletter
\providecommand \@ifxundefined [1]{%
 \@ifx{#1\undefined}
}%
\providecommand \@ifnum [1]{%
 \ifnum #1\expandafter \@firstoftwo
 \else \expandafter \@secondoftwo
 \fi
}%
\providecommand \@ifx [1]{%
 \ifx #1\expandafter \@firstoftwo
 \else \expandafter \@secondoftwo
 \fi
}%
\providecommand \natexlab [1]{#1}%
\providecommand \enquote  [1]{``#1''}%
\providecommand \bibnamefont  [1]{#1}%
\providecommand \bibfnamefont [1]{#1}%
\providecommand \citenamefont [1]{#1}%
\providecommand \href@noop [0]{\@secondoftwo}%
\providecommand \href [0]{\begingroup \@sanitize@url \@href}%
\providecommand \@href[1]{\@@startlink{#1}\@@href}%
\providecommand \@@href[1]{\endgroup#1\@@endlink}%
\providecommand \@sanitize@url [0]{\catcode `\\12\catcode `\$12\catcode
  `\&12\catcode `\#12\catcode `\^12\catcode `\_12\catcode `\%12\relax}%
\providecommand \@@startlink[1]{}%
\providecommand \@@endlink[0]{}%
\providecommand \url  [0]{\begingroup\@sanitize@url \@url }%
\providecommand \@url [1]{\endgroup\@href {#1}{\urlprefix }}%
\providecommand \urlprefix  [0]{URL }%
\providecommand \Eprint [0]{\href }%
\providecommand \doibase [0]{https://doi.org/}%
\providecommand \selectlanguage [0]{\@gobble}%
\providecommand \bibinfo  [0]{\@secondoftwo}%
\providecommand \bibfield  [0]{\@secondoftwo}%
\providecommand \translation [1]{[#1]}%
\providecommand \BibitemOpen [0]{}%
\providecommand \bibitemStop [0]{}%
\providecommand \bibitemNoStop [0]{.\EOS\space}%
\providecommand \EOS [0]{\spacefactor3000\relax}%
\providecommand \BibitemShut  [1]{\csname bibitem#1\endcsname}%
\let\auto@bib@innerbib\@empty
\bibitem [{\citenamefont {{Uzan}}(2011)}]{UzanCst2011}%
  \BibitemOpen
  \bibfield  {author} {\bibinfo {author} {\bibfnamefont {J.-P.}\ \bibnamefont
  {{Uzan}}},\ }\bibfield  {title} {\bibinfo {title} {{Varying Constants,
  Gravitation and Cosmology}},\ }\href {https://doi.org/10.12942/lrr-2011-2}
  {\bibfield  {journal} {\bibinfo  {journal} {Living Reviews in Relativity}\
  }\textbf {\bibinfo {volume} {14}},\ \bibinfo {eid} {2} (\bibinfo {year}
  {2011})},\ \Eprint {https://arxiv.org/abs/1009.5514} {arXiv:1009.5514
  [astro-ph.CO]} \BibitemShut {NoStop}%
\bibitem [{\citenamefont {{Di Casola}}\ \emph {et~al.}(2015)\citenamefont {{Di
  Casola}}, \citenamefont {{Liberati}},\ and\ \citenamefont {{Sonego}}}]{EPs}%
  \BibitemOpen
  \bibfield  {author} {\bibinfo {author} {\bibfnamefont {E.}~\bibnamefont {{Di
  Casola}}}, \bibinfo {author} {\bibfnamefont {S.}~\bibnamefont {{Liberati}}},\
  and\ \bibinfo {author} {\bibfnamefont {S.}~\bibnamefont {{Sonego}}},\
  }\bibfield  {title} {\bibinfo {title} {{Nonequivalence of equivalence
  principles}},\ }\href {https://doi.org/10.1119/1.4895342} {\bibfield
  {journal} {\bibinfo  {journal} {American Journal of Physics}\ }\textbf
  {\bibinfo {volume} {83}},\ \bibinfo {pages} {39} (\bibinfo {year} {2015})},\
  \Eprint {https://arxiv.org/abs/1310.7426} {arXiv:1310.7426 [gr-qc]}
  \BibitemShut {NoStop}%
\bibitem [{\citenamefont {{Will}}(2014)}]{Will2014}%
  \BibitemOpen
  \bibfield  {author} {\bibinfo {author} {\bibfnamefont {C.~M.}\ \bibnamefont
  {{Will}}},\ }\bibfield  {title} {\bibinfo {title} {{The Confrontation between
  General Relativity and Experiment}},\ }\href
  {https://doi.org/10.12942/lrr-2014-4} {\bibfield  {journal} {\bibinfo
  {journal} {Living Reviews in Relativity}\ }\textbf {\bibinfo {volume} {17}},\
  \bibinfo {eid} {4} (\bibinfo {year} {2014})},\ \Eprint
  {https://arxiv.org/abs/1403.7377} {arXiv:1403.7377 [gr-qc]} \BibitemShut
  {NoStop}%
\bibitem [{\citenamefont {{Will}}(2017)}]{Will2017}%
  \BibitemOpen
  \bibfield  {author} {\bibinfo {author} {\bibfnamefont {C.~M.}\ \bibnamefont
  {{Will}}},\ }\href@noop {} {\emph {\bibinfo {title} {{Theory and Experiment
  in Gravitational Physics (2nd edition)}}}}\ (\bibinfo {year}
  {2017})\BibitemShut {NoStop}%
\bibitem [{\citenamefont {{Dvali}}\ and\ \citenamefont
  {{Zaldarriaga}}(2002)}]{fifthforcealpha}%
  \BibitemOpen
  \bibfield  {author} {\bibinfo {author} {\bibfnamefont {G.}~\bibnamefont
  {{Dvali}}}\ and\ \bibinfo {author} {\bibfnamefont {M.}~\bibnamefont
  {{Zaldarriaga}}},\ }\bibfield  {title} {\bibinfo {title} {{Changing
  {\ensuremath{\alpha}} with Time: Implications for Fifth-Force-Type
  Experiments and Quintessence}},\ }\href
  {https://doi.org/10.1103/PhysRevLett.88.091303} {\bibfield  {journal}
  {\bibinfo  {journal} {\prl}\ }\textbf {\bibinfo {volume} {88}},\ \bibinfo
  {eid} {091303} (\bibinfo {year} {2002})},\ \Eprint
  {https://arxiv.org/abs/hep-ph/0108217} {arXiv:hep-ph/0108217 [hep-ph]}
  \BibitemShut {NoStop}%
\bibitem [{\citenamefont {{Martins}}(2017)}]{Martins2017review}%
  \BibitemOpen
  \bibfield  {author} {\bibinfo {author} {\bibfnamefont {C.~J.~A.~P.}\
  \bibnamefont {{Martins}}},\ }\bibfield  {title} {\bibinfo {title} {{The
  status of varying constants: a review of the physics, searches and
  implications}},\ }\href {https://doi.org/10.1088/1361-6633/aa860e} {\bibfield
   {journal} {\bibinfo  {journal} {Reports on Progress in Physics}\ }\textbf
  {\bibinfo {volume} {80}},\ \bibinfo {eid} {126902} (\bibinfo {year}
  {2017})},\ \Eprint {https://arxiv.org/abs/1709.02923} {arXiv:1709.02923
  [astro-ph.CO]} \BibitemShut {NoStop}%
\bibitem [{\citenamefont {{Martins}}(2002)}]{Cosmowithvarconst2002}%
  \BibitemOpen
  \bibfield  {author} {\bibinfo {author} {\bibfnamefont {C.~J.~A.~P.}\
  \bibnamefont {{Martins}}},\ }\bibfield  {title} {\bibinfo {title} {{Cosmology
  with varying constants}},\ }\href {https://doi.org/10.1098/rsta.2002.1087}
  {\bibfield  {journal} {\bibinfo  {journal} {Philosophical Transactions of the
  Royal Society of London Series A}\ }\textbf {\bibinfo {volume} {360}},\
  \bibinfo {pages} {2681} (\bibinfo {year} {2002})},\ \Eprint
  {https://arxiv.org/abs/astro-ph/0205504} {arXiv:astro-ph/0205504 [astro-ph]}
  \BibitemShut {NoStop}%
\bibitem [{\citenamefont {{Damour}}\ and\ \citenamefont
  {{Donoghue}}(2010)}]{damourEEPscalar}%
  \BibitemOpen
  \bibfield  {author} {\bibinfo {author} {\bibfnamefont {T.}~\bibnamefont
  {{Damour}}}\ and\ \bibinfo {author} {\bibfnamefont {J.~F.}\ \bibnamefont
  {{Donoghue}}},\ }\bibfield  {title} {\bibinfo {title} {{FAST TRACK
  COMMUNICATION: Phenomenology of the equivalence principle with light
  scalars}},\ }\href {https://doi.org/10.1088/0264-9381/27/20/202001}
  {\bibfield  {journal} {\bibinfo  {journal} {Classical and Quantum Gravity}\
  }\textbf {\bibinfo {volume} {27}},\ \bibinfo {eid} {202001} (\bibinfo {year}
  {2010})},\ \Eprint {https://arxiv.org/abs/1007.2790} {arXiv:1007.2790
  [gr-qc]} \BibitemShut {NoStop}%
\bibitem [{\citenamefont {{Martins}}\ \emph
  {et~al.}(2022{\natexlab{a}})\citenamefont {{Martins}}, \citenamefont
  {{Ferreira}},\ and\ \citenamefont {{Marto}}}]{alphacosmography}%
  \BibitemOpen
  \bibfield  {author} {\bibinfo {author} {\bibfnamefont {C.~J.~A.~P.}\
  \bibnamefont {{Martins}}}, \bibinfo {author} {\bibfnamefont {F.~P.~S.~A.}\
  \bibnamefont {{Ferreira}}},\ and\ \bibinfo {author} {\bibfnamefont {P.~V.}\
  \bibnamefont {{Marto}}},\ }\bibfield  {title} {\bibinfo {title} {{Varying
  fine-structure constant cosmography}},\ }\href
  {https://doi.org/10.1016/j.physletb.2022.137002} {\bibfield  {journal}
  {\bibinfo  {journal} {Physics Letters B}\ }\textbf {\bibinfo {volume}
  {827}},\ \bibinfo {eid} {137002} (\bibinfo {year} {2022}{\natexlab{a}})},\
  \Eprint {https://arxiv.org/abs/2203.02781} {arXiv:2203.02781 [astro-ph.CO]}
  \BibitemShut {NoStop}%
\bibitem [{\citenamefont {{Bekenstein}}(1982)}]{bekensteinOriginal}%
  \BibitemOpen
  \bibfield  {author} {\bibinfo {author} {\bibfnamefont {J.~D.}\ \bibnamefont
  {{Bekenstein}}},\ }\bibfield  {title} {\bibinfo {title} {{Fine-structure
  constant: Is it really a constant?}},\ }\href
  {https://doi.org/10.1103/PhysRevD.25.1527} {\bibfield  {journal} {\bibinfo
  {journal} {\prd}\ }\textbf {\bibinfo {volume} {25}},\ \bibinfo {pages} {1527}
  (\bibinfo {year} {1982})}\BibitemShut {NoStop}%
\bibitem [{\citenamefont {{Bekenstein}}(2002)}]{Bekenstein-2}%
  \BibitemOpen
  \bibfield  {author} {\bibinfo {author} {\bibfnamefont {J.~D.}\ \bibnamefont
  {{Bekenstein}}},\ }\bibfield  {title} {\bibinfo {title} {{Fine-structure
  constant variability, equivalence principle, and cosmology}},\ }\href
  {https://doi.org/10.1103/PhysRevD.66.123514} {\bibfield  {journal} {\bibinfo
  {journal} {\prd}\ }\textbf {\bibinfo {volume} {66}},\ \bibinfo {eid} {123514}
  (\bibinfo {year} {2002})},\ \Eprint {https://arxiv.org/abs/gr-qc/0208081}
  {arXiv:gr-qc/0208081 [gr-qc]} \BibitemShut {NoStop}%
\bibitem [{\citenamefont {{Sandvik}}\ \emph {et~al.}(2002)\citenamefont
  {{Sandvik}}, \citenamefont {{Barrow}},\ and\ \citenamefont
  {{Magueijo}}}]{Sandvik2002}%
  \BibitemOpen
  \bibfield  {author} {\bibinfo {author} {\bibfnamefont {H.~B.}\ \bibnamefont
  {{Sandvik}}}, \bibinfo {author} {\bibfnamefont {J.~D.}\ \bibnamefont
  {{Barrow}}},\ and\ \bibinfo {author} {\bibfnamefont {J.}~\bibnamefont
  {{Magueijo}}},\ }\bibfield  {title} {\bibinfo {title} {{A Simple Cosmology
  with a Varying Fine Structure Constant}},\ }\href
  {https://doi.org/10.1103/PhysRevLett.88.031302} {\bibfield  {journal}
  {\bibinfo  {journal} {\prl}\ }\textbf {\bibinfo {volume} {88}},\ \bibinfo
  {eid} {031302} (\bibinfo {year} {2002})},\ \Eprint
  {https://arxiv.org/abs/astro-ph/0107512} {arXiv:astro-ph/0107512 [astro-ph]}
  \BibitemShut {NoStop}%
\bibitem [{\citenamefont {{Leal}}\ \emph {et~al.}(2014)\citenamefont {{Leal}},
  \citenamefont {{Martins}},\ and\ \citenamefont {{Ventura}}}]{Leal2014}%
  \BibitemOpen
  \bibfield  {author} {\bibinfo {author} {\bibfnamefont {P.~M.~M.}\
  \bibnamefont {{Leal}}}, \bibinfo {author} {\bibfnamefont {C.~J.~A.~P.}\
  \bibnamefont {{Martins}}},\ and\ \bibinfo {author} {\bibfnamefont {L.~B.}\
  \bibnamefont {{Ventura}}},\ }\bibfield  {title} {\bibinfo {title}
  {{Fine-structure constant constraints on Bekenstein-type models}},\ }\href
  {https://doi.org/10.1103/PhysRevD.90.027305} {\bibfield  {journal} {\bibinfo
  {journal} {\prd}\ }\textbf {\bibinfo {volume} {90}},\ \bibinfo {eid} {027305}
  (\bibinfo {year} {2014})},\ \Eprint {https://arxiv.org/abs/1407.4099}
  {arXiv:1407.4099 [astro-ph.CO]} \BibitemShut {NoStop}%
\bibitem [{\citenamefont {{Leite}}\ and\ \citenamefont
  {{Martins}}(2016)}]{Leite2016}%
  \BibitemOpen
  \bibfield  {author} {\bibinfo {author} {\bibfnamefont {A.~C.~O.}\
  \bibnamefont {{Leite}}}\ and\ \bibinfo {author} {\bibfnamefont {C.~J.~A.~P.}\
  \bibnamefont {{Martins}}},\ }\bibfield  {title} {\bibinfo {title} {{Current
  and future constraints on Bekenstein-type models for varying couplings}},\
  }\href {https://doi.org/10.1103/PhysRevD.94.023503} {\bibfield  {journal}
  {\bibinfo  {journal} {\prd}\ }\textbf {\bibinfo {volume} {94}},\ \bibinfo
  {eid} {023503} (\bibinfo {year} {2016})},\ \Eprint
  {https://arxiv.org/abs/1607.01677} {arXiv:1607.01677 [astro-ph.CO]}
  \BibitemShut {NoStop}%
\bibitem [{\citenamefont {{Martins}}\ \emph
  {et~al.}(2022{\natexlab{b}})\citenamefont {{Martins}}, \citenamefont
  {{Cristiani}}, \citenamefont {{Cupani}}, \citenamefont {{D'Odorico}},
  \citenamefont {{G{\'e}nova Santos}}, \citenamefont {{Leite}}, \citenamefont
  {{Marques}}, \citenamefont {{Milakovi{\'c}}}, \citenamefont {{Molaro}},
  \citenamefont {{Murphy}}, \citenamefont {{Nunes}}, \citenamefont {{Schmidt}},
  \citenamefont {{Adibekyan}}, \citenamefont {{Alibert}}, \citenamefont {{Di
  Marcantonio}}, \citenamefont {{Gonz{\'a}lez Hern{\'a}ndez}}, \citenamefont
  {{M{\'e}gevand}}, \citenamefont {{Palle}}, \citenamefont {{Pepe}},
  \citenamefont {{Santos}}, \citenamefont {{Sousa}}, \citenamefont
  {{Sozzetti}}, \citenamefont {{Su{\'a}rez Mascare{\~n}o}},\ and\ \citenamefont
  {{Zapatero Osorio}}}]{MartinsEspresso2022}%
  \BibitemOpen
  \bibfield  {author} {\bibinfo {author} {\bibfnamefont {C.~J.~A.~P.}\
  \bibnamefont {{Martins}}}, \bibinfo {author} {\bibfnamefont {S.}~\bibnamefont
  {{Cristiani}}}, \bibinfo {author} {\bibfnamefont {G.}~\bibnamefont
  {{Cupani}}}, \bibinfo {author} {\bibfnamefont {V.}~\bibnamefont
  {{D'Odorico}}}, \bibinfo {author} {\bibfnamefont {R.}~\bibnamefont
  {{G{\'e}nova Santos}}}, \bibinfo {author} {\bibfnamefont {A.~C.~O.}\
  \bibnamefont {{Leite}}}, \bibinfo {author} {\bibfnamefont {C.~M.~J.}\
  \bibnamefont {{Marques}}}, \bibinfo {author} {\bibfnamefont {D.}~\bibnamefont
  {{Milakovi{\'c}}}}, \bibinfo {author} {\bibfnamefont {P.}~\bibnamefont
  {{Molaro}}}, \bibinfo {author} {\bibfnamefont {M.~T.}\ \bibnamefont
  {{Murphy}}}, \bibinfo {author} {\bibfnamefont {N.~J.}\ \bibnamefont
  {{Nunes}}}, \bibinfo {author} {\bibfnamefont {T.~M.}\ \bibnamefont
  {{Schmidt}}}, \bibinfo {author} {\bibfnamefont {V.}~\bibnamefont
  {{Adibekyan}}}, \bibinfo {author} {\bibfnamefont {Y.}~\bibnamefont
  {{Alibert}}}, \bibinfo {author} {\bibfnamefont {P.}~\bibnamefont {{Di
  Marcantonio}}}, \bibinfo {author} {\bibfnamefont {J.~I.}\ \bibnamefont
  {{Gonz{\'a}lez Hern{\'a}ndez}}}, \bibinfo {author} {\bibfnamefont
  {D.}~\bibnamefont {{M{\'e}gevand}}}, \bibinfo {author} {\bibfnamefont
  {E.}~\bibnamefont {{Palle}}}, \bibinfo {author} {\bibfnamefont {F.~A.}\
  \bibnamefont {{Pepe}}}, \bibinfo {author} {\bibfnamefont {N.~C.}\
  \bibnamefont {{Santos}}}, \bibinfo {author} {\bibfnamefont {S.~G.}\
  \bibnamefont {{Sousa}}}, \bibinfo {author} {\bibfnamefont {A.}~\bibnamefont
  {{Sozzetti}}}, \bibinfo {author} {\bibfnamefont {A.}~\bibnamefont
  {{Su{\'a}rez Mascare{\~n}o}}},\ and\ \bibinfo {author} {\bibfnamefont
  {M.~R.}\ \bibnamefont {{Zapatero Osorio}}},\ }\bibfield  {title} {\bibinfo
  {title} {{Fundamental physics with ESPRESSO: Constraints on Bekenstein and
  dark energy models from astrophysical and local probes}},\ }\href@noop {}
  {\bibfield  {journal} {\bibinfo  {journal} {arXiv e-prints}\ ,\ \bibinfo
  {eid} {arXiv:2205.13848}} (\bibinfo {year} {2022}{\natexlab{b}})},\ \Eprint
  {https://arxiv.org/abs/2205.13848} {arXiv:2205.13848 [astro-ph.CO]}
  \BibitemShut {NoStop}%
\bibitem [{\citenamefont {{Olive}}\ and\ \citenamefont
  {{Pospelov}}(2002)}]{Olive2002}%
  \BibitemOpen
  \bibfield  {author} {\bibinfo {author} {\bibfnamefont {K.~A.}\ \bibnamefont
  {{Olive}}}\ and\ \bibinfo {author} {\bibfnamefont {M.}~\bibnamefont
  {{Pospelov}}},\ }\bibfield  {title} {\bibinfo {title} {{Evolution of the fine
  structure constant driven by dark matter and the cosmological constant}},\
  }\href {https://doi.org/10.1103/PhysRevD.65.085044} {\bibfield  {journal}
  {\bibinfo  {journal} {\prd}\ }\textbf {\bibinfo {volume} {65}},\ \bibinfo
  {eid} {085044} (\bibinfo {year} {2002})},\ \Eprint
  {https://arxiv.org/abs/hep-ph/0110377} {arXiv:hep-ph/0110377 [hep-ph]}
  \BibitemShut {NoStop}%
\bibitem [{\citenamefont {{Alves}}\ \emph {et~al.}(2018)\citenamefont
  {{Alves}}, \citenamefont {{Leite}}, \citenamefont {{Martins}}, \citenamefont
  {{Silva}}, \citenamefont {{Berge}},\ and\ \citenamefont
  {{Silva}}}]{Alvesmartins2018}%
  \BibitemOpen
  \bibfield  {author} {\bibinfo {author} {\bibfnamefont {C.~S.}\ \bibnamefont
  {{Alves}}}, \bibinfo {author} {\bibfnamefont {A.~C.~O.}\ \bibnamefont
  {{Leite}}}, \bibinfo {author} {\bibfnamefont {C.~J.~A.~P.}\ \bibnamefont
  {{Martins}}}, \bibinfo {author} {\bibfnamefont {T.~A.}\ \bibnamefont
  {{Silva}}}, \bibinfo {author} {\bibfnamefont {S.~A.}\ \bibnamefont
  {{Berge}}},\ and\ \bibinfo {author} {\bibfnamefont {B.~S.~A.}\ \bibnamefont
  {{Silva}}},\ }\bibfield  {title} {\bibinfo {title} {{Current and future
  constraints on extended Bekenstein-type models for a varying fine-structure
  constant}},\ }\href {https://doi.org/10.1103/PhysRevD.97.023522} {\bibfield
  {journal} {\bibinfo  {journal} {\prd}\ }\textbf {\bibinfo {volume} {97}},\
  \bibinfo {eid} {023522} (\bibinfo {year} {2018})},\ \Eprint
  {https://arxiv.org/abs/1801.08089} {arXiv:1801.08089 [astro-ph.CO]}
  \BibitemShut {NoStop}%
\bibitem [{\citenamefont {Lesgourgues}(2011)}]{Lesgourgues:2011CLASS}%
  \BibitemOpen
  \bibfield  {author} {\bibinfo {author} {\bibfnamefont {J.}~\bibnamefont
  {Lesgourgues}},\ }\href@noop {} {\bibinfo {title} {The cosmic linear
  anisotropy solving system (class) i: Overview}} (\bibinfo {year} {2011}),\
  \Eprint {https://arxiv.org/abs/1104.2932} {arXiv:1104.2932 [astro-ph.IM]}
  \BibitemShut {NoStop}%
\bibitem [{\citenamefont {Audren}\ \emph {et~al.}(2013)\citenamefont {Audren},
  \citenamefont {Lesgourgues}, \citenamefont {Benabed},\ and\ \citenamefont
  {Prunet}}]{Audren:2012wb}%
  \BibitemOpen
  \bibfield  {author} {\bibinfo {author} {\bibfnamefont {B.}~\bibnamefont
  {Audren}}, \bibinfo {author} {\bibfnamefont {J.}~\bibnamefont {Lesgourgues}},
  \bibinfo {author} {\bibfnamefont {K.}~\bibnamefont {Benabed}},\ and\ \bibinfo
  {author} {\bibfnamefont {S.}~\bibnamefont {Prunet}},\ }\bibfield  {title}
  {\bibinfo {title} {{Conservative Constraints on Early Cosmology: an
  illustration of the Monte Python cosmological parameter inference code}},\
  }\href {https://doi.org/10.1088/1475-7516/2013/02/001} {\bibfield  {journal}
  {\bibinfo  {journal} {JCAP}\ }\textbf {\bibinfo {volume} {1302}},\ \bibinfo
  {pages} {001}},\ \Eprint {https://arxiv.org/abs/1210.7183} {arXiv:1210.7183
  [astro-ph.CO]} \BibitemShut {NoStop}%
\bibitem [{\citenamefont {{Brinckmann}}\ and\ \citenamefont
  {{Lesgourgues}}(2019)}]{Brinckmann:2018cvx}%
  \BibitemOpen
  \bibfield  {author} {\bibinfo {author} {\bibfnamefont {T.}~\bibnamefont
  {{Brinckmann}}}\ and\ \bibinfo {author} {\bibfnamefont {J.}~\bibnamefont
  {{Lesgourgues}}},\ }\bibfield  {title} {\bibinfo {title} {{MontePython 3:
  Boosted MCMC sampler and other features}},\ }\href
  {https://doi.org/10.1016/j.dark.2018.100260} {\bibfield  {journal} {\bibinfo
  {journal} {Physics of the Dark Universe}\ }\textbf {\bibinfo {volume} {24}},\
  \bibinfo {eid} {100260} (\bibinfo {year} {2019})},\ \Eprint
  {https://arxiv.org/abs/1804.07261} {arXiv:1804.07261 [astro-ph.CO]}
  \BibitemShut {NoStop}%
\bibitem [{\citenamefont {Zyla}\ \emph {et~al.}(2020)\citenamefont {Zyla} \emph
  {et~al.}}]{ParticleDataGroup:2020ssz}%
  \BibitemOpen
  \bibfield  {author} {\bibinfo {author} {\bibfnamefont {P.~A.}\ \bibnamefont
  {Zyla}} \emph {et~al.} (\bibinfo {collaboration} {Particle Data Group}),\
  }\bibfield  {title} {\bibinfo {title} {{Review of Particle Physics}},\ }\href
  {https://doi.org/10.1093/ptep/ptaa104} {\bibfield  {journal} {\bibinfo
  {journal} {PTEP}\ }\textbf {\bibinfo {volume} {2020}},\ \bibinfo {pages}
  {083C01} (\bibinfo {year} {2020})}\BibitemShut {NoStop}%
\bibitem [{\citenamefont {{Riess}}\ \emph {et~al.}(2018)\citenamefont
  {{Riess}}, \citenamefont {{Rodney}}, \citenamefont {{Scolnic}}, \citenamefont
  {{Shafer}}, \citenamefont {{Strolger}}, \citenamefont {{Ferguson}},
  \citenamefont {{Postman}}, \citenamefont {{Graur}}, \citenamefont {{Maoz}},
  \citenamefont {{Jha}}, \citenamefont {{Mobasher}}, \citenamefont
  {{Casertano}}, \citenamefont {{Hayden}}, \citenamefont {{Molino}},
  \citenamefont {{Hjorth}}, \citenamefont {{Garnavich}}, \citenamefont
  {{Jones}}, \citenamefont {{Kirshner}}, \citenamefont {{Koekemoer}},
  \citenamefont {{Grogin}}, \citenamefont {{Brammer}}, \citenamefont
  {{Hemmati}}, \citenamefont {{Dickinson}}, \citenamefont {{Challis}},
  \citenamefont {{Wolff}}, \citenamefont {{Clubb}}, \citenamefont
  {{Filippenko}}, \citenamefont {{Nayyeri}}, \citenamefont {{U}}, \citenamefont
  {{Koo}}, \citenamefont {{Faber}}, \citenamefont {{Kocevski}}, \citenamefont
  {{Bradley}},\ and\ \citenamefont {{Coe}}}]{SNPantheon}%
  \BibitemOpen
  \bibfield  {author} {\bibinfo {author} {\bibfnamefont {A.~G.}\ \bibnamefont
  {{Riess}}}, \bibinfo {author} {\bibfnamefont {S.~A.}\ \bibnamefont
  {{Rodney}}}, \bibinfo {author} {\bibfnamefont {D.~M.}\ \bibnamefont
  {{Scolnic}}}, \bibinfo {author} {\bibfnamefont {D.~L.}\ \bibnamefont
  {{Shafer}}}, \bibinfo {author} {\bibfnamefont {L.-G.}\ \bibnamefont
  {{Strolger}}}, \bibinfo {author} {\bibfnamefont {H.~C.}\ \bibnamefont
  {{Ferguson}}}, \bibinfo {author} {\bibfnamefont {M.}~\bibnamefont
  {{Postman}}}, \bibinfo {author} {\bibfnamefont {O.}~\bibnamefont {{Graur}}},
  \bibinfo {author} {\bibfnamefont {D.}~\bibnamefont {{Maoz}}}, \bibinfo
  {author} {\bibfnamefont {S.~W.}\ \bibnamefont {{Jha}}}, \bibinfo {author}
  {\bibfnamefont {B.}~\bibnamefont {{Mobasher}}}, \bibinfo {author}
  {\bibfnamefont {S.}~\bibnamefont {{Casertano}}}, \bibinfo {author}
  {\bibfnamefont {B.}~\bibnamefont {{Hayden}}}, \bibinfo {author}
  {\bibfnamefont {A.}~\bibnamefont {{Molino}}}, \bibinfo {author}
  {\bibfnamefont {J.}~\bibnamefont {{Hjorth}}}, \bibinfo {author}
  {\bibfnamefont {P.~M.}\ \bibnamefont {{Garnavich}}}, \bibinfo {author}
  {\bibfnamefont {D.~O.}\ \bibnamefont {{Jones}}}, \bibinfo {author}
  {\bibfnamefont {R.~P.}\ \bibnamefont {{Kirshner}}}, \bibinfo {author}
  {\bibfnamefont {A.~M.}\ \bibnamefont {{Koekemoer}}}, \bibinfo {author}
  {\bibfnamefont {N.~A.}\ \bibnamefont {{Grogin}}}, \bibinfo {author}
  {\bibfnamefont {G.}~\bibnamefont {{Brammer}}}, \bibinfo {author}
  {\bibfnamefont {S.}~\bibnamefont {{Hemmati}}}, \bibinfo {author}
  {\bibfnamefont {M.}~\bibnamefont {{Dickinson}}}, \bibinfo {author}
  {\bibfnamefont {P.~M.}\ \bibnamefont {{Challis}}}, \bibinfo {author}
  {\bibfnamefont {S.}~\bibnamefont {{Wolff}}}, \bibinfo {author} {\bibfnamefont
  {K.~I.}\ \bibnamefont {{Clubb}}}, \bibinfo {author} {\bibfnamefont {A.~V.}\
  \bibnamefont {{Filippenko}}}, \bibinfo {author} {\bibfnamefont
  {H.}~\bibnamefont {{Nayyeri}}}, \bibinfo {author} {\bibfnamefont
  {V.}~\bibnamefont {{U}}}, \bibinfo {author} {\bibfnamefont {D.~C.}\
  \bibnamefont {{Koo}}}, \bibinfo {author} {\bibfnamefont {S.~M.}\ \bibnamefont
  {{Faber}}}, \bibinfo {author} {\bibfnamefont {D.}~\bibnamefont {{Kocevski}}},
  \bibinfo {author} {\bibfnamefont {L.}~\bibnamefont {{Bradley}}},\ and\
  \bibinfo {author} {\bibfnamefont {D.}~\bibnamefont {{Coe}}},\ }\bibfield
  {title} {\bibinfo {title} {{Type Ia Supernova Distances at Redshift $>$1.5
  from the Hubble Space Telescope Multi-cycle Treasury Programs: The Early
  Expansion Rate}},\ }\href {https://doi.org/10.3847/1538-4357/aaa5a9}
  {\bibfield  {journal} {\bibinfo  {journal} {\apj}\ }\textbf {\bibinfo
  {volume} {853}},\ \bibinfo {eid} {126} (\bibinfo {year} {2018})},\ \Eprint
  {https://arxiv.org/abs/1710.00844} {arXiv:1710.00844 [astro-ph.CO]}
  \BibitemShut {NoStop}%
\bibitem [{\citenamefont {collaboration}(2017)}]{DR12}%
  \BibitemOpen
  \bibfield  {author} {\bibinfo {author} {\bibfnamefont {T.~B.}\ \bibnamefont
  {collaboration}},\ }\bibfield  {title} {\bibinfo {title} {{The clustering of
  galaxies in the completed SDSS-III Baryon Oscillation Spectroscopic Survey:
  cosmological analysis of the DR12 galaxy sample}},\ }\href
  {https://doi.org/10.1093/mnras/stx721} {\bibfield  {journal} {\bibinfo
  {journal} {\mnras}\ }\textbf {\bibinfo {volume} {470}},\ \bibinfo {pages}
  {2617} (\bibinfo {year} {2017})},\ \Eprint {https://arxiv.org/abs/1607.03155}
  {arXiv:1607.03155 [astro-ph.CO]} \BibitemShut {NoStop}%
\bibitem [{\citenamefont {{Moresco}}\ \emph {et~al.}(2016)\citenamefont
  {{Moresco}}, \citenamefont {{Pozzetti}}, \citenamefont {{Cimatti}},
  \citenamefont {{Jimenez}}, \citenamefont {{Maraston}}, \citenamefont
  {{Verde}}, \citenamefont {{Thomas}}, \citenamefont {{Citro}}, \citenamefont
  {{Tojeiro}},\ and\ \citenamefont {{Wilkinson}}}]{CC2016}%
  \BibitemOpen
  \bibfield  {author} {\bibinfo {author} {\bibfnamefont {M.}~\bibnamefont
  {{Moresco}}}, \bibinfo {author} {\bibfnamefont {L.}~\bibnamefont
  {{Pozzetti}}}, \bibinfo {author} {\bibfnamefont {A.}~\bibnamefont
  {{Cimatti}}}, \bibinfo {author} {\bibfnamefont {R.}~\bibnamefont
  {{Jimenez}}}, \bibinfo {author} {\bibfnamefont {C.}~\bibnamefont
  {{Maraston}}}, \bibinfo {author} {\bibfnamefont {L.}~\bibnamefont {{Verde}}},
  \bibinfo {author} {\bibfnamefont {D.}~\bibnamefont {{Thomas}}}, \bibinfo
  {author} {\bibfnamefont {A.}~\bibnamefont {{Citro}}}, \bibinfo {author}
  {\bibfnamefont {R.}~\bibnamefont {{Tojeiro}}},\ and\ \bibinfo {author}
  {\bibfnamefont {D.}~\bibnamefont {{Wilkinson}}},\ }\bibfield  {title}
  {\bibinfo {title} {{A 6\% measurement of the Hubble parameter at
  z\raisebox{-0.5ex}\textasciitilde0.45: direct evidence of the epoch of cosmic
  re-acceleration}},\ }\href {https://doi.org/10.1088/1475-7516/2016/05/014}
  {\bibfield  {journal} {\bibinfo  {journal} {\jcap}\ }\textbf {\bibinfo
  {volume} {2016}},\ \bibinfo {eid} {014} (\bibinfo {year} {2016})},\ \Eprint
  {https://arxiv.org/abs/1601.01701} {arXiv:1601.01701 [astro-ph.CO]}
  \BibitemShut {NoStop}%
\bibitem [{\citenamefont {Aghanim}\ \emph {et~al.}(2020)\citenamefont
  {Aghanim}, \citenamefont {Akrami}, \citenamefont {Arroja}, \citenamefont
  {Ashdown}, \citenamefont {Aumont}, \citenamefont {Baccigalupi}, \citenamefont
  {Ballardini}, \citenamefont {Banday}, \citenamefont {Barreiro},\ and\
  \citenamefont {et~al.}}]{Planck2018}%
  \BibitemOpen
  \bibfield  {author} {\bibinfo {author} {\bibfnamefont {N.}~\bibnamefont
  {Aghanim}}, \bibinfo {author} {\bibfnamefont {Y.}~\bibnamefont {Akrami}},
  \bibinfo {author} {\bibfnamefont {F.}~\bibnamefont {Arroja}}, \bibinfo
  {author} {\bibfnamefont {M.}~\bibnamefont {Ashdown}}, \bibinfo {author}
  {\bibfnamefont {J.}~\bibnamefont {Aumont}}, \bibinfo {author} {\bibfnamefont
  {C.}~\bibnamefont {Baccigalupi}}, \bibinfo {author} {\bibfnamefont
  {M.}~\bibnamefont {Ballardini}}, \bibinfo {author} {\bibfnamefont {A.~J.}\
  \bibnamefont {Banday}}, \bibinfo {author} {\bibfnamefont {R.~B.}\
  \bibnamefont {Barreiro}},\ and\ \bibinfo {author} {\bibnamefont {et~al.}},\
  }\bibfield  {title} {\bibinfo {title} {Planck2018 results},\ }\href
  {https://doi.org/10.1051/0004-6361/201833880} {\bibfield  {journal} {\bibinfo
   {journal} {Astronomy $\&$ Astrophysics}\ }\textbf {\bibinfo {volume}
  {641}},\ \bibinfo {pages} {A1} (\bibinfo {year} {2020})}\BibitemShut
  {NoStop}%
\bibitem [{\citenamefont {{Planck Collaboration}}(2020)}]{Planck_lkl}%
  \BibitemOpen
  \bibfield  {author} {\bibinfo {author} {\bibnamefont {{Planck
  Collaboration}}},\ }\bibfield  {title} {\bibinfo {title} {{Planck 2018
  results. V. CMB power spectra and likelihoods}},\ }\href
  {https://doi.org/10.1051/0004-6361/201936386} {\bibfield  {journal} {\bibinfo
   {journal} {\aap}\ }\textbf {\bibinfo {volume} {641}},\ \bibinfo {eid} {A5}
  (\bibinfo {year} {2020})},\ \Eprint {https://arxiv.org/abs/1907.12875}
  {arXiv:1907.12875 [astro-ph.CO]} \BibitemShut {NoStop}%
\bibitem [{\citenamefont {Murphy}\ and\ \citenamefont
  {Cooksey}(2017)}]{alphaSubaru}%
  \BibitemOpen
  \bibfield  {author} {\bibinfo {author} {\bibfnamefont {M.~T.}\ \bibnamefont
  {Murphy}}\ and\ \bibinfo {author} {\bibfnamefont {K.~L.}\ \bibnamefont
  {Cooksey}},\ }\bibfield  {title} {\bibinfo {title} {{Subaru Telescope limits
  on cosmological variations in the fine-structure constant}},\ }\href
  {https://doi.org/10.1093/mnras/stx1949} {\bibfield  {journal} {\bibinfo
  {journal} {Monthly Notices of the Royal Astronomical Society}\ }\textbf
  {\bibinfo {volume} {471}},\ \bibinfo {pages} {4930} (\bibinfo {year}
  {2017})},\ \Eprint
  {https://arxiv.org/abs/https://academic.oup.com/mnras/article-pdf/471/4/4930/19650209/stx1949.pdf}
  {https://academic.oup.com/mnras/article-pdf/471/4/4930/19650209/stx1949.pdf}
  \BibitemShut {NoStop}%
\bibitem [{\citenamefont {{Webb}}\ \emph {et~al.}(2011)\citenamefont {{Webb}},
  \citenamefont {{King}}, \citenamefont {{Murphy}}, \citenamefont {{Flambaum}},
  \citenamefont {{Carswell}},\ and\ \citenamefont {{Bainbridge}}}]{alphaWebb}%
  \BibitemOpen
  \bibfield  {author} {\bibinfo {author} {\bibfnamefont {J.~K.}\ \bibnamefont
  {{Webb}}}, \bibinfo {author} {\bibfnamefont {J.~A.}\ \bibnamefont {{King}}},
  \bibinfo {author} {\bibfnamefont {M.~T.}\ \bibnamefont {{Murphy}}}, \bibinfo
  {author} {\bibfnamefont {V.~V.}\ \bibnamefont {{Flambaum}}}, \bibinfo
  {author} {\bibfnamefont {R.~F.}\ \bibnamefont {{Carswell}}},\ and\ \bibinfo
  {author} {\bibfnamefont {M.~B.}\ \bibnamefont {{Bainbridge}}},\ }\bibfield
  {title} {\bibinfo {title} {{Indications of a Spatial Variation of the Fine
  Structure Constant}},\ }\href
  {https://doi.org/10.1103/PhysRevLett.107.191101} {\bibfield  {journal}
  {\bibinfo  {journal} {\prl}\ }\textbf {\bibinfo {volume} {107}},\ \bibinfo
  {eid} {191101} (\bibinfo {year} {2011})},\ \Eprint
  {https://arxiv.org/abs/1008.3907} {arXiv:1008.3907 [astro-ph.CO]}
  \BibitemShut {NoStop}%
\bibitem [{\citenamefont {{Murphy}}\ \emph {et~al.}(2022)\citenamefont
  {{Murphy}}, \citenamefont {{Molaro}}, \citenamefont {{Leite}}, \citenamefont
  {{Cupani}}, \citenamefont {{Cristiani}}, \citenamefont {{D'Odorico}},
  \citenamefont {{G{\'e}nova Santos}}, \citenamefont {{Martins}}, \citenamefont
  {{Milakovi{\'c}}}, \citenamefont {{Nunes}}, \citenamefont {{Schmidt}},
  \citenamefont {{Pepe}}, \citenamefont {{Rebolo}}, \citenamefont {{Santos}},
  \citenamefont {{Sousa}}, \citenamefont {{Zapatero Osorio}}, \citenamefont
  {{Amate}}, \citenamefont {{Adibekyan}}, \citenamefont {{Alibert}},
  \citenamefont {{Prieto}}, \citenamefont {{Baldini}}, \citenamefont {{Benz}},
  \citenamefont {{Bouchy}}, \citenamefont {{Cabral}}, \citenamefont {{Dekker}},
  \citenamefont {{Di Marcantonio}}, \citenamefont {{Ehrenreich}}, \citenamefont
  {{Figueira}}, \citenamefont {{Gonz{\'a}lez Hern{\'a}ndez}}, \citenamefont
  {{Landoni}}, \citenamefont {{Lovis}}, \citenamefont {{Lo Curto}},
  \citenamefont {{Manescau}}, \citenamefont {{M{\'e}gevand}}, \citenamefont
  {{Mehner}}, \citenamefont {{Micela}}, \citenamefont {{Pasquini}},
  \citenamefont {{Poretti}}, \citenamefont {{Riva}}, \citenamefont
  {{Sozzetti}}, \citenamefont {{Mascare{\~n}o}}, \citenamefont {{Udry}},\ and\
  \citenamefont {{Zerbi}}}]{alphaespresso}%
  \BibitemOpen
  \bibfield  {author} {\bibinfo {author} {\bibfnamefont {M.~T.}\ \bibnamefont
  {{Murphy}}}, \bibinfo {author} {\bibfnamefont {P.}~\bibnamefont {{Molaro}}},
  \bibinfo {author} {\bibfnamefont {A.~C.~O.}\ \bibnamefont {{Leite}}},
  \bibinfo {author} {\bibfnamefont {G.}~\bibnamefont {{Cupani}}}, \bibinfo
  {author} {\bibfnamefont {S.}~\bibnamefont {{Cristiani}}}, \bibinfo {author}
  {\bibfnamefont {V.}~\bibnamefont {{D'Odorico}}}, \bibinfo {author}
  {\bibfnamefont {R.}~\bibnamefont {{G{\'e}nova Santos}}}, \bibinfo {author}
  {\bibfnamefont {C.~J.~A.~P.}\ \bibnamefont {{Martins}}}, \bibinfo {author}
  {\bibfnamefont {D.}~\bibnamefont {{Milakovi{\'c}}}}, \bibinfo {author}
  {\bibfnamefont {N.~J.}\ \bibnamefont {{Nunes}}}, \bibinfo {author}
  {\bibfnamefont {T.~M.}\ \bibnamefont {{Schmidt}}}, \bibinfo {author}
  {\bibfnamefont {F.~A.}\ \bibnamefont {{Pepe}}}, \bibinfo {author}
  {\bibfnamefont {R.}~\bibnamefont {{Rebolo}}}, \bibinfo {author}
  {\bibfnamefont {N.~C.}\ \bibnamefont {{Santos}}}, \bibinfo {author}
  {\bibfnamefont {S.~G.}\ \bibnamefont {{Sousa}}}, \bibinfo {author}
  {\bibfnamefont {M.-R.}\ \bibnamefont {{Zapatero Osorio}}}, \bibinfo {author}
  {\bibfnamefont {M.}~\bibnamefont {{Amate}}}, \bibinfo {author} {\bibfnamefont
  {V.}~\bibnamefont {{Adibekyan}}}, \bibinfo {author} {\bibfnamefont
  {Y.}~\bibnamefont {{Alibert}}}, \bibinfo {author} {\bibfnamefont {C.~A.}\
  \bibnamefont {{Prieto}}}, \bibinfo {author} {\bibfnamefont {V.}~\bibnamefont
  {{Baldini}}}, \bibinfo {author} {\bibfnamefont {W.}~\bibnamefont {{Benz}}},
  \bibinfo {author} {\bibfnamefont {F.}~\bibnamefont {{Bouchy}}}, \bibinfo
  {author} {\bibfnamefont {A.}~\bibnamefont {{Cabral}}}, \bibinfo {author}
  {\bibfnamefont {H.}~\bibnamefont {{Dekker}}}, \bibinfo {author}
  {\bibfnamefont {P.}~\bibnamefont {{Di Marcantonio}}}, \bibinfo {author}
  {\bibfnamefont {D.}~\bibnamefont {{Ehrenreich}}}, \bibinfo {author}
  {\bibfnamefont {P.}~\bibnamefont {{Figueira}}}, \bibinfo {author}
  {\bibfnamefont {J.~I.}\ \bibnamefont {{Gonz{\'a}lez Hern{\'a}ndez}}},
  \bibinfo {author} {\bibfnamefont {M.}~\bibnamefont {{Landoni}}}, \bibinfo
  {author} {\bibfnamefont {C.}~\bibnamefont {{Lovis}}}, \bibinfo {author}
  {\bibfnamefont {G.}~\bibnamefont {{Lo Curto}}}, \bibinfo {author}
  {\bibfnamefont {A.}~\bibnamefont {{Manescau}}}, \bibinfo {author}
  {\bibfnamefont {D.}~\bibnamefont {{M{\'e}gevand}}}, \bibinfo {author}
  {\bibfnamefont {A.}~\bibnamefont {{Mehner}}}, \bibinfo {author}
  {\bibfnamefont {G.}~\bibnamefont {{Micela}}}, \bibinfo {author}
  {\bibfnamefont {L.}~\bibnamefont {{Pasquini}}}, \bibinfo {author}
  {\bibfnamefont {E.}~\bibnamefont {{Poretti}}}, \bibinfo {author}
  {\bibfnamefont {M.}~\bibnamefont {{Riva}}}, \bibinfo {author} {\bibfnamefont
  {A.}~\bibnamefont {{Sozzetti}}}, \bibinfo {author} {\bibfnamefont {A.~S.}\
  \bibnamefont {{Mascare{\~n}o}}}, \bibinfo {author} {\bibfnamefont
  {S.}~\bibnamefont {{Udry}}},\ and\ \bibinfo {author} {\bibfnamefont
  {F.}~\bibnamefont {{Zerbi}}},\ }\bibfield  {title} {\bibinfo {title}
  {{Fundamental physics with ESPRESSO: Precise limit on variations in the
  fine-structure constant towards the bright quasar HE
  0515{\ensuremath{-}}4414}},\ }\href
  {https://doi.org/10.1051/0004-6361/202142257} {\bibfield  {journal} {\bibinfo
   {journal} {\aap}\ }\textbf {\bibinfo {volume} {658}},\ \bibinfo {eid} {A123}
  (\bibinfo {year} {2022})},\ \Eprint {https://arxiv.org/abs/2112.05819}
  {arXiv:2112.05819 [astro-ph.CO]} \BibitemShut {NoStop}%
\bibitem [{\citenamefont {{Petrov}}\ \emph {et~al.}(2006)\citenamefont
  {{Petrov}}, \citenamefont {{Nazarov}}, \citenamefont {{Onegin}},
  \citenamefont {{Petrov}},\ and\ \citenamefont {{Sakhnovsky}}}]{Oklo}%
  \BibitemOpen
  \bibfield  {author} {\bibinfo {author} {\bibfnamefont {Y.~V.}\ \bibnamefont
  {{Petrov}}}, \bibinfo {author} {\bibfnamefont {A.~I.}\ \bibnamefont
  {{Nazarov}}}, \bibinfo {author} {\bibfnamefont {M.~S.}\ \bibnamefont
  {{Onegin}}}, \bibinfo {author} {\bibfnamefont {V.~Y.}\ \bibnamefont
  {{Petrov}}},\ and\ \bibinfo {author} {\bibfnamefont {E.~G.}\ \bibnamefont
  {{Sakhnovsky}}},\ }\bibfield  {title} {\bibinfo {title} {{Natural nuclear
  reactor at Oklo and variation of fundamental constants: Computation of
  neutronics of a fresh core}},\ }\href
  {https://doi.org/10.1103/PhysRevC.74.064610} {\bibfield  {journal} {\bibinfo
  {journal} {\prc}\ }\textbf {\bibinfo {volume} {74}},\ \bibinfo {eid} {064610}
  (\bibinfo {year} {2006})},\ \Eprint {https://arxiv.org/abs/hep-ph/0506186}
  {arXiv:hep-ph/0506186 [hep-ph]} \BibitemShut {NoStop}%
\bibitem [{\citenamefont {{Lange}}\ \emph {et~al.}(2021)\citenamefont
  {{Lange}}, \citenamefont {{Huntemann}}, \citenamefont {{Rahm}}, \citenamefont
  {{Sanner}}, \citenamefont {{Shao}}, \citenamefont {{Lipphardt}},
  \citenamefont {{Tamm}}, \citenamefont {{Weyers}},\ and\ \citenamefont
  {{Peik}}}]{atomicclock}%
  \BibitemOpen
  \bibfield  {author} {\bibinfo {author} {\bibfnamefont {R.}~\bibnamefont
  {{Lange}}}, \bibinfo {author} {\bibfnamefont {N.}~\bibnamefont
  {{Huntemann}}}, \bibinfo {author} {\bibfnamefont {J.~M.}\ \bibnamefont
  {{Rahm}}}, \bibinfo {author} {\bibfnamefont {C.}~\bibnamefont {{Sanner}}},
  \bibinfo {author} {\bibfnamefont {H.}~\bibnamefont {{Shao}}}, \bibinfo
  {author} {\bibfnamefont {B.}~\bibnamefont {{Lipphardt}}}, \bibinfo {author}
  {\bibfnamefont {C.}~\bibnamefont {{Tamm}}}, \bibinfo {author} {\bibfnamefont
  {S.}~\bibnamefont {{Weyers}}},\ and\ \bibinfo {author} {\bibfnamefont
  {E.}~\bibnamefont {{Peik}}},\ }\bibfield  {title} {\bibinfo {title}
  {{Improved Limits for Violations of Local Position Invariance from Atomic
  Clock Comparisons}},\ }\href {https://doi.org/10.1103/PhysRevLett.126.011102}
  {\bibfield  {journal} {\bibinfo  {journal} {\prl}\ }\textbf {\bibinfo
  {volume} {126}},\ \bibinfo {eid} {011102} (\bibinfo {year} {2021})},\ \Eprint
  {https://arxiv.org/abs/2010.06620} {arXiv:2010.06620 [physics.atom-ph]}
  \BibitemShut {NoStop}%
\bibitem [{\citenamefont {Touboul}\ \emph {et~al.}(2022)\citenamefont
  {Touboul}, \citenamefont {M\'etris}, \citenamefont {Rodrigues}, \citenamefont
  {Berg\'e}, \citenamefont {Robert}, \citenamefont {Baghi}, \citenamefont
  {Andr\'e}, \citenamefont {Bedouet}, \citenamefont {Boulanger}, \citenamefont
  {Bremer}, \citenamefont {Carle}, \citenamefont {Chhun}, \citenamefont
  {Christophe}, \citenamefont {Cipolla}, \citenamefont {Damour}, \citenamefont
  {Danto}, \citenamefont {Demange}, \citenamefont {Dittus}, \citenamefont
  {Dhuicque}, \citenamefont {Fayet}, \citenamefont {Foulon}, \citenamefont
  {Guidotti}, \citenamefont {Hagedorn}, \citenamefont {Hardy}, \citenamefont
  {Huynh}, \citenamefont {Kayser}, \citenamefont {Lala}, \citenamefont
  {L\"ammerzahl}, \citenamefont {Lebat}, \citenamefont {Liorzou}, \citenamefont
  {List}, \citenamefont {L\"offler}, \citenamefont {Panet}, \citenamefont
  {Pernot-Borr\`as}, \citenamefont {Perraud}, \citenamefont {Pires},
  \citenamefont {Pouilloux}, \citenamefont {Prieur}, \citenamefont {Rebray},
  \citenamefont {Reynaud}, \citenamefont {Rievers}, \citenamefont {Selig},
  \citenamefont {Serron}, \citenamefont {Sumner}, \citenamefont {Tanguy},
  \citenamefont {Torresi},\ and\ \citenamefont {Visser}}]{Microscope_new}%
  \BibitemOpen
  \bibfield  {author} {\bibinfo {author} {\bibfnamefont {P.}~\bibnamefont
  {Touboul}}, \bibinfo {author} {\bibfnamefont {G.}~\bibnamefont {M\'etris}},
  \bibinfo {author} {\bibfnamefont {M.}~\bibnamefont {Rodrigues}}, \bibinfo
  {author} {\bibfnamefont {J.}~\bibnamefont {Berg\'e}}, \bibinfo {author}
  {\bibfnamefont {A.}~\bibnamefont {Robert}}, \bibinfo {author} {\bibfnamefont
  {Q.}~\bibnamefont {Baghi}}, \bibinfo {author} {\bibfnamefont
  {Y.}~\bibnamefont {Andr\'e}}, \bibinfo {author} {\bibfnamefont
  {J.}~\bibnamefont {Bedouet}}, \bibinfo {author} {\bibfnamefont
  {D.}~\bibnamefont {Boulanger}}, \bibinfo {author} {\bibfnamefont
  {S.}~\bibnamefont {Bremer}}, \bibinfo {author} {\bibfnamefont
  {P.}~\bibnamefont {Carle}}, \bibinfo {author} {\bibfnamefont
  {R.}~\bibnamefont {Chhun}}, \bibinfo {author} {\bibfnamefont
  {B.}~\bibnamefont {Christophe}}, \bibinfo {author} {\bibfnamefont
  {V.}~\bibnamefont {Cipolla}}, \bibinfo {author} {\bibfnamefont
  {T.}~\bibnamefont {Damour}}, \bibinfo {author} {\bibfnamefont
  {P.}~\bibnamefont {Danto}}, \bibinfo {author} {\bibfnamefont
  {L.}~\bibnamefont {Demange}}, \bibinfo {author} {\bibfnamefont
  {H.}~\bibnamefont {Dittus}}, \bibinfo {author} {\bibfnamefont
  {O.}~\bibnamefont {Dhuicque}}, \bibinfo {author} {\bibfnamefont
  {P.}~\bibnamefont {Fayet}}, \bibinfo {author} {\bibfnamefont
  {B.}~\bibnamefont {Foulon}}, \bibinfo {author} {\bibfnamefont {P.-Y.}\
  \bibnamefont {Guidotti}}, \bibinfo {author} {\bibfnamefont {D.}~\bibnamefont
  {Hagedorn}}, \bibinfo {author} {\bibfnamefont {E.}~\bibnamefont {Hardy}},
  \bibinfo {author} {\bibfnamefont {P.-A.}\ \bibnamefont {Huynh}}, \bibinfo
  {author} {\bibfnamefont {P.}~\bibnamefont {Kayser}}, \bibinfo {author}
  {\bibfnamefont {S.}~\bibnamefont {Lala}}, \bibinfo {author} {\bibfnamefont
  {C.}~\bibnamefont {L\"ammerzahl}}, \bibinfo {author} {\bibfnamefont
  {V.}~\bibnamefont {Lebat}}, \bibinfo {author} {\bibfnamefont {F.~m.~c.}\
  \bibnamefont {Liorzou}}, \bibinfo {author} {\bibfnamefont {M.}~\bibnamefont
  {List}}, \bibinfo {author} {\bibfnamefont {F.}~\bibnamefont {L\"offler}},
  \bibinfo {author} {\bibfnamefont {I.}~\bibnamefont {Panet}}, \bibinfo
  {author} {\bibfnamefont {M.}~\bibnamefont {Pernot-Borr\`as}}, \bibinfo
  {author} {\bibfnamefont {L.}~\bibnamefont {Perraud}}, \bibinfo {author}
  {\bibfnamefont {S.}~\bibnamefont {Pires}}, \bibinfo {author} {\bibfnamefont
  {B.}~\bibnamefont {Pouilloux}}, \bibinfo {author} {\bibfnamefont
  {P.}~\bibnamefont {Prieur}}, \bibinfo {author} {\bibfnamefont
  {A.}~\bibnamefont {Rebray}}, \bibinfo {author} {\bibfnamefont
  {S.}~\bibnamefont {Reynaud}}, \bibinfo {author} {\bibfnamefont
  {B.}~\bibnamefont {Rievers}}, \bibinfo {author} {\bibfnamefont
  {H.}~\bibnamefont {Selig}}, \bibinfo {author} {\bibfnamefont
  {L.}~\bibnamefont {Serron}}, \bibinfo {author} {\bibfnamefont
  {T.}~\bibnamefont {Sumner}}, \bibinfo {author} {\bibfnamefont
  {N.}~\bibnamefont {Tanguy}}, \bibinfo {author} {\bibfnamefont
  {P.}~\bibnamefont {Torresi}},\ and\ \bibinfo {author} {\bibfnamefont
  {P.}~\bibnamefont {Visser}},\ }\bibfield  {title} {\bibinfo {title}
  {$microscope$ mission: Final results of the test of the equivalence
  principle},\ }\href {https://doi.org/10.1103/PhysRevLett.129.121102}
  {\bibfield  {journal} {\bibinfo  {journal} {Phys. Rev. Lett.}\ }\textbf
  {\bibinfo {volume} {129}},\ \bibinfo {pages} {121102} (\bibinfo {year}
  {2022})}\BibitemShut {NoStop}%
\bibitem [{\citenamefont {{Planck Collaboration}}(2015)}]{Planckalphame2016}%
  \BibitemOpen
  \bibfield  {author} {\bibinfo {author} {\bibnamefont {{Planck
  Collaboration}}},\ }\bibfield  {title} {\bibinfo {title} {{Planck
  intermediate results. XXIV. Constraints on variations in fundamental
  constants}},\ }\href {https://doi.org/10.1051/0004-6361/201424496} {\bibfield
   {journal} {\bibinfo  {journal} {\aap}\ }\textbf {\bibinfo {volume} {580}},\
  \bibinfo {eid} {A22} (\bibinfo {year} {2015})},\ \Eprint
  {https://arxiv.org/abs/1406.7482} {arXiv:1406.7482 [astro-ph.CO]}
  \BibitemShut {NoStop}%
\bibitem [{\citenamefont {{Hart}}\ and\ \citenamefont
  {{Chluba}}(2018)}]{HartChluba}%
  \BibitemOpen
  \bibfield  {author} {\bibinfo {author} {\bibfnamefont {L.}~\bibnamefont
  {{Hart}}}\ and\ \bibinfo {author} {\bibfnamefont {J.}~\bibnamefont
  {{Chluba}}},\ }\bibfield  {title} {\bibinfo {title} {{New constraints on
  time-dependent variations of fundamental constants using Planck data}},\
  }\href {https://doi.org/10.1093/mnras/stx2783} {\bibfield  {journal}
  {\bibinfo  {journal} {\mnras}\ }\textbf {\bibinfo {volume} {474}},\ \bibinfo
  {pages} {1850} (\bibinfo {year} {2018})},\ \Eprint
  {https://arxiv.org/abs/1705.03925} {arXiv:1705.03925 [astro-ph.CO]}
  \BibitemShut {NoStop}%
\bibitem [{\citenamefont {{Lewis}}(2019)}]{getdist}%
  \BibitemOpen
  \bibfield  {author} {\bibinfo {author} {\bibfnamefont {A.}~\bibnamefont
  {{Lewis}}},\ }\bibfield  {title} {\bibinfo {title} {{GetDist: a Python
  package for analysing Monte Carlo samples}},\ }\href@noop {} {\bibfield
  {journal} {\bibinfo  {journal} {arXiv e-prints}\ ,\ \bibinfo {eid}
  {arXiv:1910.13970}} (\bibinfo {year} {2019})},\ \Eprint
  {https://arxiv.org/abs/1910.13970} {arXiv:1910.13970 [astro-ph.IM]}
  \BibitemShut {NoStop}%
\bibitem [{\citenamefont {Rosenband}\ \emph {et~al.}(2008)\citenamefont
  {Rosenband}, \citenamefont {Hume}, \citenamefont {Schmidt}, \citenamefont
  {Chou}, \citenamefont {Brusch}, \citenamefont {Lorini}, \citenamefont
  {Oskay}, \citenamefont {Drullinger}, \citenamefont {Fortier}, \citenamefont
  {Stalnaker}, \citenamefont {Diddams}, \citenamefont {Swann}, \citenamefont
  {Newbury}, \citenamefont {Itano}, \citenamefont {Wineland},\ and\
  \citenamefont {Bergquist}}]{Rosenband}%
  \BibitemOpen
  \bibfield  {author} {\bibinfo {author} {\bibfnamefont {T.}~\bibnamefont
  {Rosenband}}, \bibinfo {author} {\bibfnamefont {D.~B.}\ \bibnamefont {Hume}},
  \bibinfo {author} {\bibfnamefont {P.~O.}\ \bibnamefont {Schmidt}}, \bibinfo
  {author} {\bibfnamefont {C.~W.}\ \bibnamefont {Chou}}, \bibinfo {author}
  {\bibfnamefont {A.}~\bibnamefont {Brusch}}, \bibinfo {author} {\bibfnamefont
  {L.}~\bibnamefont {Lorini}}, \bibinfo {author} {\bibfnamefont {W.~H.}\
  \bibnamefont {Oskay}}, \bibinfo {author} {\bibfnamefont {R.~E.}\ \bibnamefont
  {Drullinger}}, \bibinfo {author} {\bibfnamefont {T.~M.}\ \bibnamefont
  {Fortier}}, \bibinfo {author} {\bibfnamefont {J.~E.}\ \bibnamefont
  {Stalnaker}}, \bibinfo {author} {\bibfnamefont {S.~A.}\ \bibnamefont
  {Diddams}}, \bibinfo {author} {\bibfnamefont {W.~C.}\ \bibnamefont {Swann}},
  \bibinfo {author} {\bibfnamefont {N.~R.}\ \bibnamefont {Newbury}}, \bibinfo
  {author} {\bibfnamefont {W.~M.}\ \bibnamefont {Itano}}, \bibinfo {author}
  {\bibfnamefont {D.~J.}\ \bibnamefont {Wineland}},\ and\ \bibinfo {author}
  {\bibfnamefont {J.~C.}\ \bibnamefont {Bergquist}},\ }\bibfield  {title}
  {\bibinfo {title} {Frequency ratio of al${}^+$ and hg${}^+$ single-ion
  optical clocks; metrology at the 17th decimal place},\ }\href
  {https://doi.org/10.1126/science.1154622} {\bibfield  {journal} {\bibinfo
  {journal} {Science}\ }\textbf {\bibinfo {volume} {319}},\ \bibinfo {pages}
  {1808} (\bibinfo {year} {2008})},\ \Eprint
  {https://arxiv.org/abs/https://www.science.org/doi/pdf/10.1126/science.1154622}
  {https://www.science.org/doi/pdf/10.1126/science.1154622} \BibitemShut
  {NoStop}%
\bibitem [{\citenamefont {{Wagner}}\ \emph {et~al.}(2012)\citenamefont
  {{Wagner}}, \citenamefont {{Schlamminger}}, \citenamefont {{Gundlach}},\ and\
  \citenamefont {{Adelberger}}}]{Eot-wash}%
  \BibitemOpen
  \bibfield  {author} {\bibinfo {author} {\bibfnamefont {T.~A.}\ \bibnamefont
  {{Wagner}}}, \bibinfo {author} {\bibfnamefont {S.}~\bibnamefont
  {{Schlamminger}}}, \bibinfo {author} {\bibfnamefont {J.~H.}\ \bibnamefont
  {{Gundlach}}},\ and\ \bibinfo {author} {\bibfnamefont {E.~G.}\ \bibnamefont
  {{Adelberger}}},\ }\bibfield  {title} {\bibinfo {title} {{Torsion-balance
  tests of the weak equivalence principle}},\ }\href
  {https://doi.org/10.1088/0264-9381/29/18/184002} {\bibfield  {journal}
  {\bibinfo  {journal} {Classical and Quantum Gravity}\ }\textbf {\bibinfo
  {volume} {29}},\ \bibinfo {eid} {184002} (\bibinfo {year} {2012})},\ \Eprint
  {https://arxiv.org/abs/1207.2442} {arXiv:1207.2442 [gr-qc]} \BibitemShut
  {NoStop}%
\bibitem [{\citenamefont {Hart}\ and\ \citenamefont
  {Chluba}(2022)}]{Hart:2021kad}%
  \BibitemOpen
  \bibfield  {author} {\bibinfo {author} {\bibfnamefont {L.}~\bibnamefont
  {Hart}}\ and\ \bibinfo {author} {\bibfnamefont {J.}~\bibnamefont {Chluba}},\
  }\bibfield  {title} {\bibinfo {title} {{Varying fundamental constants
  principal component analysis: additional hints about the Hubble tension}},\
  }\href {https://doi.org/10.1093/mnras/stab2777} {\bibfield  {journal}
  {\bibinfo  {journal} {Mon. Not. Roy. Astron. Soc.}\ }\textbf {\bibinfo
  {volume} {510}},\ \bibinfo {pages} {2206} (\bibinfo {year} {2022})},\ \Eprint
  {https://arxiv.org/abs/2107.12465} {arXiv:2107.12465 [astro-ph.CO]}
  \BibitemShut {NoStop}%
\bibitem [{\citenamefont {{Sch{\"o}neberg}}\ \emph {et~al.}(2021)\citenamefont
  {{Sch{\"o}neberg}}, \citenamefont {{Abell{\'a}n}}, \citenamefont {{P{\'e}rez
  S{\'a}nchez}}, \citenamefont {{Witte}}, \citenamefont {{Poulin}},\ and\
  \citenamefont {{Lesgourgues}}}]{H0olympics}%
  \BibitemOpen
  \bibfield  {author} {\bibinfo {author} {\bibfnamefont {N.}~\bibnamefont
  {{Sch{\"o}neberg}}}, \bibinfo {author} {\bibfnamefont {G.~F.}\ \bibnamefont
  {{Abell{\'a}n}}}, \bibinfo {author} {\bibfnamefont {A.}~\bibnamefont
  {{P{\'e}rez S{\'a}nchez}}}, \bibinfo {author} {\bibfnamefont {S.~J.}\
  \bibnamefont {{Witte}}}, \bibinfo {author} {\bibfnamefont {V.}~\bibnamefont
  {{Poulin}}},\ and\ \bibinfo {author} {\bibfnamefont {J.}~\bibnamefont
  {{Lesgourgues}}},\ }\bibfield  {title} {\bibinfo {title} {{The $H_0$
  Olympics: A fair ranking of proposed models}},\ }\href@noop {} {\bibfield
  {journal} {\bibinfo  {journal} {arXiv e-prints}\ ,\ \bibinfo {eid}
  {arXiv:2107.10291}} (\bibinfo {year} {2021})},\ \Eprint
  {https://arxiv.org/abs/2107.10291} {arXiv:2107.10291 [astro-ph.CO]}
  \BibitemShut {NoStop}%
\bibitem [{\citenamefont {Liske}\ \emph {et~al.}(2014)\citenamefont {Liske},
  \citenamefont {Bono}, \citenamefont {Cepa} \emph {et~al.}}]{HIRES}%
  \BibitemOpen
  \bibfield  {author} {\bibinfo {author} {\bibfnamefont {J.}~\bibnamefont
  {Liske}}, \bibinfo {author} {\bibfnamefont {G.}~\bibnamefont {Bono}},
  \bibinfo {author} {\bibfnamefont {J.}~\bibnamefont {Cepa}}, \emph {et~al.},\
  }\href@noop {} {\emph {\bibinfo {title} {{Top Level Requirements For
  ELT-HIRES}}}},\ \bibinfo {type} {Tech. Rep.}\ (\bibinfo  {institution}
  {Document ESO 204697 Version 1},\ \bibinfo {year} {2014})\BibitemShut
  {NoStop}%
\bibitem [{\citenamefont {Fadeev}\ \emph {et~al.}(2020)\citenamefont {Fadeev},
  \citenamefont {Berengut},\ and\ \citenamefont {Flambaum}}]{THorium}%
  \BibitemOpen
  \bibfield  {author} {\bibinfo {author} {\bibfnamefont {P.}~\bibnamefont
  {Fadeev}}, \bibinfo {author} {\bibfnamefont {J.~C.}\ \bibnamefont
  {Berengut}},\ and\ \bibinfo {author} {\bibfnamefont {V.~V.}\ \bibnamefont
  {Flambaum}},\ }\bibfield  {title} {\bibinfo {title} {{Sensitivity of
  $^{229}$Th nuclear clock transition to variation of the fine-structure
  constant}},\ }\href {https://doi.org/10.1103/PhysRevA.102.052833} {\bibfield
  {journal} {\bibinfo  {journal} {Phys. Rev. A}\ }\textbf {\bibinfo {volume}
  {102}},\ \bibinfo {pages} {052833} (\bibinfo {year} {2020})},\ \Eprint
  {https://arxiv.org/abs/2007.00408} {arXiv:2007.00408 [physics.atom-ph]}
  \BibitemShut {NoStop}%
\bibitem [{\citenamefont {{Martinelli}}\ \emph {et~al.}(2021)\citenamefont
  {{Martinelli}}, \citenamefont {{Martins}}, \citenamefont {{Nesseris}},
  \citenamefont {{Tutusaus}}, \citenamefont {{Blanchard}}, \citenamefont
  {{Camera}}, \citenamefont {{Carbone}}, \citenamefont {{Casas}}, \citenamefont
  {{Pettorino}}, \citenamefont {{Sakr}}, \citenamefont {{Yankelevich}},
  \citenamefont {{Sapone}}, \citenamefont {{Amara}}, \citenamefont
  {{Auricchio}}, \citenamefont {{Bodendorf}}, \citenamefont {{Bonino}},
  \citenamefont {{Branchini}}, \citenamefont {{Capobianco}}, \citenamefont
  {{Carretero}}, \citenamefont {{Castellano}}, \citenamefont {{Cavuoti}},
  \citenamefont {{Cimatti}}, \citenamefont {{Cledassou}}, \citenamefont
  {{Corcione}}, \citenamefont {{Costille}}, \citenamefont {{Degaudenzi}},
  \citenamefont {{Douspis}}, \citenamefont {{Dubath}}, \citenamefont
  {{Dusini}}, \citenamefont {{Ealet}}, \citenamefont {{Ferriol}}, \citenamefont
  {{Frailis}}, \citenamefont {{Franceschi}}, \citenamefont {{Garilli}},
  \citenamefont {{Giocoli}}, \citenamefont {{Grazian}}, \citenamefont
  {{Grupp}}, \citenamefont {{Haugan}}, \citenamefont {{Holmes}}, \citenamefont
  {{Hormuth}}, \citenamefont {{Jahnke}}, \citenamefont {{Kiessling}},
  \citenamefont {{K{\"u}mmel}}, \citenamefont {{Kunz}}, \citenamefont
  {{Kurki-Suonio}}, \citenamefont {{Ligori}}, \citenamefont {{Lilje}},
  \citenamefont {{Lloro}}, \citenamefont {{Mansutti}}, \citenamefont
  {{Marggraf}}, \citenamefont {{Markovic}}, \citenamefont {{Massey}},
  \citenamefont {{Meneghetti}}, \citenamefont {{Meylan}}, \citenamefont
  {{Moscardini}}, \citenamefont {{Niemi}}, \citenamefont {{Padilla}},
  \citenamefont {{Paltani}}, \citenamefont {{Pasian}}, \citenamefont
  {{Pedersen}}, \citenamefont {{Pires}}, \citenamefont {{Poncet}},
  \citenamefont {{Popa}}, \citenamefont {{Raison}}, \citenamefont {{Rebolo}},
  \citenamefont {{Rhodes}}, \citenamefont {{Roncarelli}}, \citenamefont
  {{Rossetti}}, \citenamefont {{Saglia}}, \citenamefont {{Secroun}},
  \citenamefont {{Seidel}}, \citenamefont {{Serrano}}, \citenamefont
  {{Sirignano}}, \citenamefont {{Sirri}}, \citenamefont {{Starck}},
  \citenamefont {{Tavagnacco}}, \citenamefont {{Taylor}}, \citenamefont
  {{Tereno}}, \citenamefont {{Toledo-Moreo}}, \citenamefont {{Valenziano}},
  \citenamefont {{Wang}}, \citenamefont {{Zamorani}}, \citenamefont
  {{Zoubian}}, \citenamefont {{Baldi}}, \citenamefont {{Brescia}},
  \citenamefont {{Congedo}}, \citenamefont {{Conversi}}, \citenamefont
  {{Copin}}, \citenamefont {{Fabbian}}, \citenamefont {{Farinelli}},
  \citenamefont {{Medinaceli}}, \citenamefont {{Mei}}, \citenamefont
  {{Polenta}}, \citenamefont {{Romelli}},\ and\ \citenamefont
  {{Vassallo}}}]{Euclid2021}%
  \BibitemOpen
  \bibfield  {author} {\bibinfo {author} {\bibfnamefont {M.}~\bibnamefont
  {{Martinelli}}}, \bibinfo {author} {\bibfnamefont {C.~J.~A.~P.}\ \bibnamefont
  {{Martins}}}, \bibinfo {author} {\bibfnamefont {S.}~\bibnamefont
  {{Nesseris}}}, \bibinfo {author} {\bibfnamefont {I.}~\bibnamefont
  {{Tutusaus}}}, \bibinfo {author} {\bibfnamefont {A.}~\bibnamefont
  {{Blanchard}}}, \bibinfo {author} {\bibfnamefont {S.}~\bibnamefont
  {{Camera}}}, \bibinfo {author} {\bibfnamefont {C.}~\bibnamefont {{Carbone}}},
  \bibinfo {author} {\bibfnamefont {S.}~\bibnamefont {{Casas}}}, \bibinfo
  {author} {\bibfnamefont {V.}~\bibnamefont {{Pettorino}}}, \bibinfo {author}
  {\bibfnamefont {Z.}~\bibnamefont {{Sakr}}}, \bibinfo {author} {\bibfnamefont
  {V.}~\bibnamefont {{Yankelevich}}}, \bibinfo {author} {\bibfnamefont
  {D.}~\bibnamefont {{Sapone}}}, \bibinfo {author} {\bibfnamefont
  {A.}~\bibnamefont {{Amara}}}, \bibinfo {author} {\bibfnamefont
  {N.}~\bibnamefont {{Auricchio}}}, \bibinfo {author} {\bibfnamefont
  {C.}~\bibnamefont {{Bodendorf}}}, \bibinfo {author} {\bibfnamefont
  {D.}~\bibnamefont {{Bonino}}}, \bibinfo {author} {\bibfnamefont
  {E.}~\bibnamefont {{Branchini}}}, \bibinfo {author} {\bibfnamefont
  {V.}~\bibnamefont {{Capobianco}}}, \bibinfo {author} {\bibfnamefont
  {J.}~\bibnamefont {{Carretero}}}, \bibinfo {author} {\bibfnamefont
  {M.}~\bibnamefont {{Castellano}}}, \bibinfo {author} {\bibfnamefont
  {S.}~\bibnamefont {{Cavuoti}}}, \bibinfo {author} {\bibfnamefont
  {A.}~\bibnamefont {{Cimatti}}}, \bibinfo {author} {\bibfnamefont
  {R.}~\bibnamefont {{Cledassou}}}, \bibinfo {author} {\bibfnamefont
  {L.}~\bibnamefont {{Corcione}}}, \bibinfo {author} {\bibfnamefont
  {A.}~\bibnamefont {{Costille}}}, \bibinfo {author} {\bibfnamefont
  {H.}~\bibnamefont {{Degaudenzi}}}, \bibinfo {author} {\bibfnamefont
  {M.}~\bibnamefont {{Douspis}}}, \bibinfo {author} {\bibfnamefont
  {F.}~\bibnamefont {{Dubath}}}, \bibinfo {author} {\bibfnamefont
  {S.}~\bibnamefont {{Dusini}}}, \bibinfo {author} {\bibfnamefont
  {A.}~\bibnamefont {{Ealet}}}, \bibinfo {author} {\bibfnamefont
  {S.}~\bibnamefont {{Ferriol}}}, \bibinfo {author} {\bibfnamefont
  {M.}~\bibnamefont {{Frailis}}}, \bibinfo {author} {\bibfnamefont
  {E.}~\bibnamefont {{Franceschi}}}, \bibinfo {author} {\bibfnamefont
  {B.}~\bibnamefont {{Garilli}}}, \bibinfo {author} {\bibfnamefont
  {C.}~\bibnamefont {{Giocoli}}}, \bibinfo {author} {\bibfnamefont
  {A.}~\bibnamefont {{Grazian}}}, \bibinfo {author} {\bibfnamefont
  {F.}~\bibnamefont {{Grupp}}}, \bibinfo {author} {\bibfnamefont {S.~V.~H.}\
  \bibnamefont {{Haugan}}}, \bibinfo {author} {\bibfnamefont {W.}~\bibnamefont
  {{Holmes}}}, \bibinfo {author} {\bibfnamefont {F.}~\bibnamefont {{Hormuth}}},
  \bibinfo {author} {\bibfnamefont {K.}~\bibnamefont {{Jahnke}}}, \bibinfo
  {author} {\bibfnamefont {A.}~\bibnamefont {{Kiessling}}}, \bibinfo {author}
  {\bibfnamefont {M.}~\bibnamefont {{K{\"u}mmel}}}, \bibinfo {author}
  {\bibfnamefont {M.}~\bibnamefont {{Kunz}}}, \bibinfo {author} {\bibfnamefont
  {H.}~\bibnamefont {{Kurki-Suonio}}}, \bibinfo {author} {\bibfnamefont
  {S.}~\bibnamefont {{Ligori}}}, \bibinfo {author} {\bibfnamefont {P.~B.}\
  \bibnamefont {{Lilje}}}, \bibinfo {author} {\bibfnamefont {I.}~\bibnamefont
  {{Lloro}}}, \bibinfo {author} {\bibfnamefont {O.}~\bibnamefont {{Mansutti}}},
  \bibinfo {author} {\bibfnamefont {O.}~\bibnamefont {{Marggraf}}}, \bibinfo
  {author} {\bibfnamefont {K.}~\bibnamefont {{Markovic}}}, \bibinfo {author}
  {\bibfnamefont {R.}~\bibnamefont {{Massey}}}, \bibinfo {author}
  {\bibfnamefont {M.}~\bibnamefont {{Meneghetti}}}, \bibinfo {author}
  {\bibfnamefont {G.}~\bibnamefont {{Meylan}}}, \bibinfo {author}
  {\bibfnamefont {L.}~\bibnamefont {{Moscardini}}}, \bibinfo {author}
  {\bibfnamefont {S.~M.}\ \bibnamefont {{Niemi}}}, \bibinfo {author}
  {\bibfnamefont {C.}~\bibnamefont {{Padilla}}}, \bibinfo {author}
  {\bibfnamefont {S.}~\bibnamefont {{Paltani}}}, \bibinfo {author}
  {\bibfnamefont {F.}~\bibnamefont {{Pasian}}}, \bibinfo {author}
  {\bibfnamefont {K.}~\bibnamefont {{Pedersen}}}, \bibinfo {author}
  {\bibfnamefont {S.}~\bibnamefont {{Pires}}}, \bibinfo {author} {\bibfnamefont
  {M.}~\bibnamefont {{Poncet}}}, \bibinfo {author} {\bibfnamefont
  {L.}~\bibnamefont {{Popa}}}, \bibinfo {author} {\bibfnamefont
  {F.}~\bibnamefont {{Raison}}}, \bibinfo {author} {\bibfnamefont
  {R.}~\bibnamefont {{Rebolo}}}, \bibinfo {author} {\bibfnamefont
  {J.}~\bibnamefont {{Rhodes}}}, \bibinfo {author} {\bibfnamefont
  {M.}~\bibnamefont {{Roncarelli}}}, \bibinfo {author} {\bibfnamefont
  {E.}~\bibnamefont {{Rossetti}}}, \bibinfo {author} {\bibfnamefont
  {R.}~\bibnamefont {{Saglia}}}, \bibinfo {author} {\bibfnamefont
  {A.}~\bibnamefont {{Secroun}}}, \bibinfo {author} {\bibfnamefont
  {G.}~\bibnamefont {{Seidel}}}, \bibinfo {author} {\bibfnamefont
  {S.}~\bibnamefont {{Serrano}}}, \bibinfo {author} {\bibfnamefont
  {C.}~\bibnamefont {{Sirignano}}}, \bibinfo {author} {\bibfnamefont
  {G.}~\bibnamefont {{Sirri}}}, \bibinfo {author} {\bibfnamefont {J.~L.}\
  \bibnamefont {{Starck}}}, \bibinfo {author} {\bibfnamefont {D.}~\bibnamefont
  {{Tavagnacco}}}, \bibinfo {author} {\bibfnamefont {A.~N.}\ \bibnamefont
  {{Taylor}}}, \bibinfo {author} {\bibfnamefont {I.}~\bibnamefont {{Tereno}}},
  \bibinfo {author} {\bibfnamefont {R.}~\bibnamefont {{Toledo-Moreo}}},
  \bibinfo {author} {\bibfnamefont {L.}~\bibnamefont {{Valenziano}}}, \bibinfo
  {author} {\bibfnamefont {Y.}~\bibnamefont {{Wang}}}, \bibinfo {author}
  {\bibfnamefont {G.}~\bibnamefont {{Zamorani}}}, \bibinfo {author}
  {\bibfnamefont {J.}~\bibnamefont {{Zoubian}}}, \bibinfo {author}
  {\bibfnamefont {M.}~\bibnamefont {{Baldi}}}, \bibinfo {author} {\bibfnamefont
  {M.}~\bibnamefont {{Brescia}}}, \bibinfo {author} {\bibfnamefont
  {G.}~\bibnamefont {{Congedo}}}, \bibinfo {author} {\bibfnamefont
  {L.}~\bibnamefont {{Conversi}}}, \bibinfo {author} {\bibfnamefont
  {Y.}~\bibnamefont {{Copin}}}, \bibinfo {author} {\bibfnamefont
  {G.}~\bibnamefont {{Fabbian}}}, \bibinfo {author} {\bibfnamefont
  {R.}~\bibnamefont {{Farinelli}}}, \bibinfo {author} {\bibfnamefont
  {E.}~\bibnamefont {{Medinaceli}}}, \bibinfo {author} {\bibfnamefont
  {S.}~\bibnamefont {{Mei}}}, \bibinfo {author} {\bibfnamefont
  {G.}~\bibnamefont {{Polenta}}}, \bibinfo {author} {\bibfnamefont
  {E.}~\bibnamefont {{Romelli}}},\ and\ \bibinfo {author} {\bibfnamefont
  {T.}~\bibnamefont {{Vassallo}}},\ }\bibfield  {title} {\bibinfo {title}
  {{Euclid: Constraining dark energy coupled to electromagnetism using
  astrophysical and laboratory data}},\ }\href
  {https://doi.org/10.1051/0004-6361/202141353} {\bibfield  {journal} {\bibinfo
   {journal} {\aap}\ }\textbf {\bibinfo {volume} {654}},\ \bibinfo {eid} {A148}
  (\bibinfo {year} {2021})},\ \Eprint {https://arxiv.org/abs/2105.09746}
  {arXiv:2105.09746 [astro-ph.CO]} \BibitemShut {NoStop}%
\bibitem [{\citenamefont {{DESI Collaboration}}(2016)}]{Desi2016}%
  \BibitemOpen
  \bibfield  {author} {\bibinfo {author} {\bibnamefont {{DESI
  Collaboration}}},\ }\bibfield  {title} {\bibinfo {title} {{The DESI
  Experiment Part I: Science,Targeting, and Survey Design}},\ }\href@noop {}
  {\bibfield  {journal} {\bibinfo  {journal} {arXiv e-prints}\ ,\ \bibinfo
  {eid} {arXiv:1611.00036}} (\bibinfo {year} {2016})},\ \Eprint
  {https://arxiv.org/abs/1611.00036} {arXiv:1611.00036 [astro-ph.IM]}
  \BibitemShut {NoStop}%
\bibitem [{\citenamefont {{Aiola}}\ \emph {et~al.}(2020)\citenamefont
  {{Aiola}}, \citenamefont {{Calabrese}}, \citenamefont {{Maurin}},
  \citenamefont {{Naess}}, \citenamefont {{Schmitt}}, \citenamefont
  {{Abitbol}}, \citenamefont {{Addison}}, \citenamefont {{Ade}}, \citenamefont
  {{Alonso}}, \citenamefont {{Amiri}}, \citenamefont {{Amodeo}}, \citenamefont
  {{Angile}}, \citenamefont {{Austermann}}, \citenamefont {{Baildon}},
  \citenamefont {{Battaglia}}, \citenamefont {{Beall}}, \citenamefont {{Bean}},
  \citenamefont {{Becker}}, \citenamefont {{Bond}}, \citenamefont {{Bruno}},
  \citenamefont {{Calafut}}, \citenamefont {{Campusano}}, \citenamefont
  {{Carrero}}, \citenamefont {{Chesmore}}, \citenamefont {{Cho}}, \citenamefont
  {{Choi}}, \citenamefont {{Clark}}, \citenamefont {{Cothard}}, \citenamefont
  {{Crichton}}, \citenamefont {{Crowley}}, \citenamefont {{Darwish}},
  \citenamefont {{Datta}}, \citenamefont {{Denison}}, \citenamefont {{Devlin}},
  \citenamefont {{Duell}}, \citenamefont {{Duff}}, \citenamefont
  {{Duivenvoorden}}, \citenamefont {{Dunkley}}, \citenamefont {{D{\"u}nner}},
  \citenamefont {{Essinger-Hileman}}, \citenamefont {{Fankhanel}},
  \citenamefont {{Ferraro}}, \citenamefont {{Fox}}, \citenamefont {{Fuzia}},
  \citenamefont {{Gallardo}}, \citenamefont {{Gluscevic}}, \citenamefont
  {{Golec}}, \citenamefont {{Grace}}, \citenamefont {{Gralla}}, \citenamefont
  {{Guan}}, \citenamefont {{Hall}}, \citenamefont {{Halpern}}, \citenamefont
  {{Han}}, \citenamefont {{Hargrave}}, \citenamefont {{Hasselfield}},
  \citenamefont {{Helton}}, \citenamefont {{Henderson}}, \citenamefont
  {{Hensley}}, \citenamefont {{Hill}}, \citenamefont {{Hilton}}, \citenamefont
  {{Hilton}}, \citenamefont {{Hincks}}, \citenamefont {{Hlo{\v{z}}ek}},
  \citenamefont {{Ho}}, \citenamefont {{Hubmayr}}, \citenamefont
  {{Huffenberger}}, \citenamefont {{Hughes}}, \citenamefont {{Infante}},
  \citenamefont {{Irwin}}, \citenamefont {{Jackson}}, \citenamefont {{Klein}},
  \citenamefont {{Knowles}}, \citenamefont {{Koopman}}, \citenamefont
  {{Kosowsky}}, \citenamefont {{Lakey}}, \citenamefont {{Li}}, \citenamefont
  {{Li}}, \citenamefont {{Li}}, \citenamefont {{Lokken}}, \citenamefont
  {{Louis}}, \citenamefont {{Lungu}}, \citenamefont {{MacInnis}}, \citenamefont
  {{Madhavacheril}}, \citenamefont {{Maldonado}}, \citenamefont
  {{Mallaby-Kay}}, \citenamefont {{Marsden}}, \citenamefont {{McMahon}},
  \citenamefont {{Menanteau}}, \citenamefont {{Moodley}}, \citenamefont
  {{Morton}}, \citenamefont {{Namikawa}}, \citenamefont {{Nati}}, \citenamefont
  {{Newburgh}}, \citenamefont {{Nibarger}}, \citenamefont {{Nicola}},
  \citenamefont {{Niemack}}, \citenamefont {{Nolta}}, \citenamefont
  {{Orlowski-Sherer}}, \citenamefont {{Page}}, \citenamefont {{Pappas}},
  \citenamefont {{Partridge}}, \citenamefont {{Phakathi}}, \citenamefont
  {{Pisano}}, \citenamefont {{Prince}}, \citenamefont {{Puddu}}, \citenamefont
  {{Qu}}, \citenamefont {{Rivera}}, \citenamefont {{Robertson}}, \citenamefont
  {{Rojas}}, \citenamefont {{Salatino}}, \citenamefont {{Schaan}},
  \citenamefont {{Schillaci}}, \citenamefont {{Sehgal}}, \citenamefont
  {{Sherwin}}, \citenamefont {{Sierra}}, \citenamefont {{Sievers}},
  \citenamefont {{Sifon}}, \citenamefont {{Sikhosana}}, \citenamefont
  {{Simon}}, \citenamefont {{Spergel}}, \citenamefont {{Staggs}}, \citenamefont
  {{Stevens}}, \citenamefont {{Storer}}, \citenamefont {{Sunder}},
  \citenamefont {{Switzer}}, \citenamefont {{Thorne}}, \citenamefont
  {{Thornton}}, \citenamefont {{Trac}}, \citenamefont {{Treu}}, \citenamefont
  {{Tucker}}, \citenamefont {{Vale}}, \citenamefont {{Van Engelen}},
  \citenamefont {{Van Lanen}}, \citenamefont {{Vavagiakis}}, \citenamefont
  {{Wagoner}}, \citenamefont {{Wang}}, \citenamefont {{Ward}}, \citenamefont
  {{Wollack}}, \citenamefont {{Xu}}, \citenamefont {{Zago}},\ and\
  \citenamefont {{Zhu}}}]{ACT}%
  \BibitemOpen
  \bibfield  {author} {\bibinfo {author} {\bibfnamefont {S.}~\bibnamefont
  {{Aiola}}}, \bibinfo {author} {\bibfnamefont {E.}~\bibnamefont
  {{Calabrese}}}, \bibinfo {author} {\bibfnamefont {L.}~\bibnamefont
  {{Maurin}}}, \bibinfo {author} {\bibfnamefont {S.}~\bibnamefont {{Naess}}},
  \bibinfo {author} {\bibfnamefont {B.~L.}\ \bibnamefont {{Schmitt}}}, \bibinfo
  {author} {\bibfnamefont {M.~H.}\ \bibnamefont {{Abitbol}}}, \bibinfo {author}
  {\bibfnamefont {G.~E.}\ \bibnamefont {{Addison}}}, \bibinfo {author}
  {\bibfnamefont {P.~A.~R.}\ \bibnamefont {{Ade}}}, \bibinfo {author}
  {\bibfnamefont {D.}~\bibnamefont {{Alonso}}}, \bibinfo {author}
  {\bibfnamefont {M.}~\bibnamefont {{Amiri}}}, \bibinfo {author} {\bibfnamefont
  {S.}~\bibnamefont {{Amodeo}}}, \bibinfo {author} {\bibfnamefont
  {E.}~\bibnamefont {{Angile}}}, \bibinfo {author} {\bibfnamefont {J.~E.}\
  \bibnamefont {{Austermann}}}, \bibinfo {author} {\bibfnamefont
  {T.}~\bibnamefont {{Baildon}}}, \bibinfo {author} {\bibfnamefont
  {N.}~\bibnamefont {{Battaglia}}}, \bibinfo {author} {\bibfnamefont {J.~A.}\
  \bibnamefont {{Beall}}}, \bibinfo {author} {\bibfnamefont {R.}~\bibnamefont
  {{Bean}}}, \bibinfo {author} {\bibfnamefont {D.~T.}\ \bibnamefont
  {{Becker}}}, \bibinfo {author} {\bibfnamefont {J.~R.}\ \bibnamefont
  {{Bond}}}, \bibinfo {author} {\bibfnamefont {S.~M.}\ \bibnamefont {{Bruno}}},
  \bibinfo {author} {\bibfnamefont {V.}~\bibnamefont {{Calafut}}}, \bibinfo
  {author} {\bibfnamefont {L.~E.}\ \bibnamefont {{Campusano}}}, \bibinfo
  {author} {\bibfnamefont {F.}~\bibnamefont {{Carrero}}}, \bibinfo {author}
  {\bibfnamefont {G.~E.}\ \bibnamefont {{Chesmore}}}, \bibinfo {author}
  {\bibfnamefont {H.-m.}\ \bibnamefont {{Cho}}}, \bibinfo {author}
  {\bibfnamefont {S.~K.}\ \bibnamefont {{Choi}}}, \bibinfo {author}
  {\bibfnamefont {S.~E.}\ \bibnamefont {{Clark}}}, \bibinfo {author}
  {\bibfnamefont {N.~F.}\ \bibnamefont {{Cothard}}}, \bibinfo {author}
  {\bibfnamefont {D.}~\bibnamefont {{Crichton}}}, \bibinfo {author}
  {\bibfnamefont {K.~T.}\ \bibnamefont {{Crowley}}}, \bibinfo {author}
  {\bibfnamefont {O.}~\bibnamefont {{Darwish}}}, \bibinfo {author}
  {\bibfnamefont {R.}~\bibnamefont {{Datta}}}, \bibinfo {author} {\bibfnamefont
  {E.~V.}\ \bibnamefont {{Denison}}}, \bibinfo {author} {\bibfnamefont {M.~J.}\
  \bibnamefont {{Devlin}}}, \bibinfo {author} {\bibfnamefont {C.~J.}\
  \bibnamefont {{Duell}}}, \bibinfo {author} {\bibfnamefont {S.~M.}\
  \bibnamefont {{Duff}}}, \bibinfo {author} {\bibfnamefont {A.~J.}\
  \bibnamefont {{Duivenvoorden}}}, \bibinfo {author} {\bibfnamefont
  {J.}~\bibnamefont {{Dunkley}}}, \bibinfo {author} {\bibfnamefont
  {R.}~\bibnamefont {{D{\"u}nner}}}, \bibinfo {author} {\bibfnamefont
  {T.}~\bibnamefont {{Essinger-Hileman}}}, \bibinfo {author} {\bibfnamefont
  {M.}~\bibnamefont {{Fankhanel}}}, \bibinfo {author} {\bibfnamefont
  {S.}~\bibnamefont {{Ferraro}}}, \bibinfo {author} {\bibfnamefont {A.~E.}\
  \bibnamefont {{Fox}}}, \bibinfo {author} {\bibfnamefont {B.}~\bibnamefont
  {{Fuzia}}}, \bibinfo {author} {\bibfnamefont {P.~A.}\ \bibnamefont
  {{Gallardo}}}, \bibinfo {author} {\bibfnamefont {V.}~\bibnamefont
  {{Gluscevic}}}, \bibinfo {author} {\bibfnamefont {J.~E.}\ \bibnamefont
  {{Golec}}}, \bibinfo {author} {\bibfnamefont {E.}~\bibnamefont {{Grace}}},
  \bibinfo {author} {\bibfnamefont {M.}~\bibnamefont {{Gralla}}}, \bibinfo
  {author} {\bibfnamefont {Y.}~\bibnamefont {{Guan}}}, \bibinfo {author}
  {\bibfnamefont {K.}~\bibnamefont {{Hall}}}, \bibinfo {author} {\bibfnamefont
  {M.}~\bibnamefont {{Halpern}}}, \bibinfo {author} {\bibfnamefont
  {D.}~\bibnamefont {{Han}}}, \bibinfo {author} {\bibfnamefont
  {P.}~\bibnamefont {{Hargrave}}}, \bibinfo {author} {\bibfnamefont
  {M.}~\bibnamefont {{Hasselfield}}}, \bibinfo {author} {\bibfnamefont {J.~M.}\
  \bibnamefont {{Helton}}}, \bibinfo {author} {\bibfnamefont {S.}~\bibnamefont
  {{Henderson}}}, \bibinfo {author} {\bibfnamefont {B.}~\bibnamefont
  {{Hensley}}}, \bibinfo {author} {\bibfnamefont {J.~C.}\ \bibnamefont
  {{Hill}}}, \bibinfo {author} {\bibfnamefont {G.~C.}\ \bibnamefont
  {{Hilton}}}, \bibinfo {author} {\bibfnamefont {M.}~\bibnamefont {{Hilton}}},
  \bibinfo {author} {\bibfnamefont {A.~D.}\ \bibnamefont {{Hincks}}}, \bibinfo
  {author} {\bibfnamefont {R.}~\bibnamefont {{Hlo{\v{z}}ek}}}, \bibinfo
  {author} {\bibfnamefont {S.-P.~P.}\ \bibnamefont {{Ho}}}, \bibinfo {author}
  {\bibfnamefont {J.}~\bibnamefont {{Hubmayr}}}, \bibinfo {author}
  {\bibfnamefont {K.~M.}\ \bibnamefont {{Huffenberger}}}, \bibinfo {author}
  {\bibfnamefont {J.~P.}\ \bibnamefont {{Hughes}}}, \bibinfo {author}
  {\bibfnamefont {L.}~\bibnamefont {{Infante}}}, \bibinfo {author}
  {\bibfnamefont {K.}~\bibnamefont {{Irwin}}}, \bibinfo {author} {\bibfnamefont
  {R.}~\bibnamefont {{Jackson}}}, \bibinfo {author} {\bibfnamefont
  {J.}~\bibnamefont {{Klein}}}, \bibinfo {author} {\bibfnamefont
  {K.}~\bibnamefont {{Knowles}}}, \bibinfo {author} {\bibfnamefont
  {B.}~\bibnamefont {{Koopman}}}, \bibinfo {author} {\bibfnamefont
  {A.}~\bibnamefont {{Kosowsky}}}, \bibinfo {author} {\bibfnamefont
  {V.}~\bibnamefont {{Lakey}}}, \bibinfo {author} {\bibfnamefont
  {D.}~\bibnamefont {{Li}}}, \bibinfo {author} {\bibfnamefont {Y.}~\bibnamefont
  {{Li}}}, \bibinfo {author} {\bibfnamefont {Z.}~\bibnamefont {{Li}}}, \bibinfo
  {author} {\bibfnamefont {M.}~\bibnamefont {{Lokken}}}, \bibinfo {author}
  {\bibfnamefont {T.}~\bibnamefont {{Louis}}}, \bibinfo {author} {\bibfnamefont
  {M.}~\bibnamefont {{Lungu}}}, \bibinfo {author} {\bibfnamefont
  {A.}~\bibnamefont {{MacInnis}}}, \bibinfo {author} {\bibfnamefont
  {M.}~\bibnamefont {{Madhavacheril}}}, \bibinfo {author} {\bibfnamefont
  {F.}~\bibnamefont {{Maldonado}}}, \bibinfo {author} {\bibfnamefont
  {M.}~\bibnamefont {{Mallaby-Kay}}}, \bibinfo {author} {\bibfnamefont
  {D.}~\bibnamefont {{Marsden}}}, \bibinfo {author} {\bibfnamefont
  {J.}~\bibnamefont {{McMahon}}}, \bibinfo {author} {\bibfnamefont
  {F.}~\bibnamefont {{Menanteau}}}, \bibinfo {author} {\bibfnamefont
  {K.}~\bibnamefont {{Moodley}}}, \bibinfo {author} {\bibfnamefont
  {T.}~\bibnamefont {{Morton}}}, \bibinfo {author} {\bibfnamefont
  {T.}~\bibnamefont {{Namikawa}}}, \bibinfo {author} {\bibfnamefont
  {F.}~\bibnamefont {{Nati}}}, \bibinfo {author} {\bibfnamefont
  {L.}~\bibnamefont {{Newburgh}}}, \bibinfo {author} {\bibfnamefont {J.~P.}\
  \bibnamefont {{Nibarger}}}, \bibinfo {author} {\bibfnamefont
  {A.}~\bibnamefont {{Nicola}}}, \bibinfo {author} {\bibfnamefont {M.~D.}\
  \bibnamefont {{Niemack}}}, \bibinfo {author} {\bibfnamefont {M.~R.}\
  \bibnamefont {{Nolta}}}, \bibinfo {author} {\bibfnamefont {J.}~\bibnamefont
  {{Orlowski-Sherer}}}, \bibinfo {author} {\bibfnamefont {L.~A.}\ \bibnamefont
  {{Page}}}, \bibinfo {author} {\bibfnamefont {C.~G.}\ \bibnamefont
  {{Pappas}}}, \bibinfo {author} {\bibfnamefont {B.}~\bibnamefont
  {{Partridge}}}, \bibinfo {author} {\bibfnamefont {P.}~\bibnamefont
  {{Phakathi}}}, \bibinfo {author} {\bibfnamefont {G.}~\bibnamefont
  {{Pisano}}}, \bibinfo {author} {\bibfnamefont {H.}~\bibnamefont {{Prince}}},
  \bibinfo {author} {\bibfnamefont {R.}~\bibnamefont {{Puddu}}}, \bibinfo
  {author} {\bibfnamefont {F.~J.}\ \bibnamefont {{Qu}}}, \bibinfo {author}
  {\bibfnamefont {J.}~\bibnamefont {{Rivera}}}, \bibinfo {author}
  {\bibfnamefont {N.}~\bibnamefont {{Robertson}}}, \bibinfo {author}
  {\bibfnamefont {F.}~\bibnamefont {{Rojas}}}, \bibinfo {author} {\bibfnamefont
  {M.}~\bibnamefont {{Salatino}}}, \bibinfo {author} {\bibfnamefont
  {E.}~\bibnamefont {{Schaan}}}, \bibinfo {author} {\bibfnamefont
  {A.}~\bibnamefont {{Schillaci}}}, \bibinfo {author} {\bibfnamefont
  {N.}~\bibnamefont {{Sehgal}}}, \bibinfo {author} {\bibfnamefont {B.~D.}\
  \bibnamefont {{Sherwin}}}, \bibinfo {author} {\bibfnamefont {C.}~\bibnamefont
  {{Sierra}}}, \bibinfo {author} {\bibfnamefont {J.}~\bibnamefont {{Sievers}}},
  \bibinfo {author} {\bibfnamefont {C.}~\bibnamefont {{Sifon}}}, \bibinfo
  {author} {\bibfnamefont {P.}~\bibnamefont {{Sikhosana}}}, \bibinfo {author}
  {\bibfnamefont {S.}~\bibnamefont {{Simon}}}, \bibinfo {author} {\bibfnamefont
  {D.~N.}\ \bibnamefont {{Spergel}}}, \bibinfo {author} {\bibfnamefont {S.~T.}\
  \bibnamefont {{Staggs}}}, \bibinfo {author} {\bibfnamefont {J.}~\bibnamefont
  {{Stevens}}}, \bibinfo {author} {\bibfnamefont {E.}~\bibnamefont {{Storer}}},
  \bibinfo {author} {\bibfnamefont {D.~D.}\ \bibnamefont {{Sunder}}}, \bibinfo
  {author} {\bibfnamefont {E.~R.}\ \bibnamefont {{Switzer}}}, \bibinfo {author}
  {\bibfnamefont {B.}~\bibnamefont {{Thorne}}}, \bibinfo {author}
  {\bibfnamefont {R.}~\bibnamefont {{Thornton}}}, \bibinfo {author}
  {\bibfnamefont {H.}~\bibnamefont {{Trac}}}, \bibinfo {author} {\bibfnamefont
  {J.}~\bibnamefont {{Treu}}}, \bibinfo {author} {\bibfnamefont
  {C.}~\bibnamefont {{Tucker}}}, \bibinfo {author} {\bibfnamefont {L.~R.}\
  \bibnamefont {{Vale}}}, \bibinfo {author} {\bibfnamefont {A.}~\bibnamefont
  {{Van Engelen}}}, \bibinfo {author} {\bibfnamefont {J.}~\bibnamefont {{Van
  Lanen}}}, \bibinfo {author} {\bibfnamefont {E.~M.}\ \bibnamefont
  {{Vavagiakis}}}, \bibinfo {author} {\bibfnamefont {K.}~\bibnamefont
  {{Wagoner}}}, \bibinfo {author} {\bibfnamefont {Y.}~\bibnamefont {{Wang}}},
  \bibinfo {author} {\bibfnamefont {J.~T.}\ \bibnamefont {{Ward}}}, \bibinfo
  {author} {\bibfnamefont {E.~J.}\ \bibnamefont {{Wollack}}}, \bibinfo {author}
  {\bibfnamefont {Z.}~\bibnamefont {{Xu}}}, \bibinfo {author} {\bibfnamefont
  {F.}~\bibnamefont {{Zago}}},\ and\ \bibinfo {author} {\bibfnamefont
  {N.}~\bibnamefont {{Zhu}}},\ }\bibfield  {title} {\bibinfo {title} {{The
  Atacama Cosmology Telescope: DR4 maps and cosmological parameters}},\ }\href
  {https://doi.org/10.1088/1475-7516/2020/12/047} {\bibfield  {journal}
  {\bibinfo  {journal} {\jcap}\ }\textbf {\bibinfo {volume} {2020}},\ \bibinfo
  {eid} {047} (\bibinfo {year} {2020})},\ \Eprint
  {https://arxiv.org/abs/2007.07288} {arXiv:2007.07288 [astro-ph.CO]}
  \BibitemShut {NoStop}%
\bibitem [{\citenamefont {{Sayre}}\ \emph {et~al.}(2020)\citenamefont
  {{Sayre}}, \citenamefont {{Reichardt}}, \citenamefont {{Henning}},
  \citenamefont {{Ade}}, \citenamefont {{Anderson}}, \citenamefont
  {{Austermann}}, \citenamefont {{Avva}}, \citenamefont {{Beall}},
  \citenamefont {{Bender}}, \citenamefont {{Benson}}, \citenamefont
  {{Bianchini}}, \citenamefont {{Bleem}}, \citenamefont {{Carlstrom}},
  \citenamefont {{Chang}}, \citenamefont {{Chaubal}}, \citenamefont {{Chiang}},
  \citenamefont {{Citron}}, \citenamefont {{Corbett Moran}}, \citenamefont
  {{Crawford}}, \citenamefont {{Crites}}, \citenamefont {{de Haan}},
  \citenamefont {{Dobbs}}, \citenamefont {{Everett}}, \citenamefont
  {{Gallicchio}}, \citenamefont {{George}}, \citenamefont {{Gilbert}},
  \citenamefont {{Gupta}}, \citenamefont {{Halverson}}, \citenamefont
  {{Harrington}}, \citenamefont {{Hilton}}, \citenamefont {{Holder}},
  \citenamefont {{Holzapfel}}, \citenamefont {{Hrubes}}, \citenamefont
  {{Huang}}, \citenamefont {{Hubmayr}}, \citenamefont {{Irwin}}, \citenamefont
  {{Knox}}, \citenamefont {{Lee}}, \citenamefont {{Li}}, \citenamefont
  {{Lowitz}}, \citenamefont {{McMahon}}, \citenamefont {{Meyer}}, \citenamefont
  {{Mocanu}}, \citenamefont {{Montgomery}}, \citenamefont {{Nadolski}},
  \citenamefont {{Natoli}}, \citenamefont {{Nibarger}}, \citenamefont
  {{Noble}}, \citenamefont {{Novosad}}, \citenamefont {{Padin}}, \citenamefont
  {{Patil}}, \citenamefont {{Pryke}}, \citenamefont {{Ruhl}}, \citenamefont
  {{Saliwanchik}}, \citenamefont {{Schaffer}}, \citenamefont {{Sievers}},
  \citenamefont {{Smecher}}, \citenamefont {{Stark}}, \citenamefont {{Tucker}},
  \citenamefont {{Vanderlinde}}, \citenamefont {{Veach}}, \citenamefont
  {{Vieira}}, \citenamefont {{Wang}}, \citenamefont {{Whitehorn}},
  \citenamefont {{Wu}}, \citenamefont {{Yefremenko}},\ and\ \citenamefont
  {{SPTpol Collaboration}}}]{SPT}%
  \BibitemOpen
  \bibfield  {author} {\bibinfo {author} {\bibfnamefont {J.~T.}\ \bibnamefont
  {{Sayre}}}, \bibinfo {author} {\bibfnamefont {C.~L.}\ \bibnamefont
  {{Reichardt}}}, \bibinfo {author} {\bibfnamefont {J.~W.}\ \bibnamefont
  {{Henning}}}, \bibinfo {author} {\bibfnamefont {P.~A.~R.}\ \bibnamefont
  {{Ade}}}, \bibinfo {author} {\bibfnamefont {A.~J.}\ \bibnamefont
  {{Anderson}}}, \bibinfo {author} {\bibfnamefont {J.~E.}\ \bibnamefont
  {{Austermann}}}, \bibinfo {author} {\bibfnamefont {J.~S.}\ \bibnamefont
  {{Avva}}}, \bibinfo {author} {\bibfnamefont {J.~A.}\ \bibnamefont {{Beall}}},
  \bibinfo {author} {\bibfnamefont {A.~N.}\ \bibnamefont {{Bender}}}, \bibinfo
  {author} {\bibfnamefont {B.~A.}\ \bibnamefont {{Benson}}}, \bibinfo {author}
  {\bibfnamefont {F.}~\bibnamefont {{Bianchini}}}, \bibinfo {author}
  {\bibfnamefont {L.~E.}\ \bibnamefont {{Bleem}}}, \bibinfo {author}
  {\bibfnamefont {J.~E.}\ \bibnamefont {{Carlstrom}}}, \bibinfo {author}
  {\bibfnamefont {C.~L.}\ \bibnamefont {{Chang}}}, \bibinfo {author}
  {\bibfnamefont {P.}~\bibnamefont {{Chaubal}}}, \bibinfo {author}
  {\bibfnamefont {H.~C.}\ \bibnamefont {{Chiang}}}, \bibinfo {author}
  {\bibfnamefont {R.}~\bibnamefont {{Citron}}}, \bibinfo {author}
  {\bibfnamefont {C.}~\bibnamefont {{Corbett Moran}}}, \bibinfo {author}
  {\bibfnamefont {T.~M.}\ \bibnamefont {{Crawford}}}, \bibinfo {author}
  {\bibfnamefont {A.~T.}\ \bibnamefont {{Crites}}}, \bibinfo {author}
  {\bibfnamefont {T.}~\bibnamefont {{de Haan}}}, \bibinfo {author}
  {\bibfnamefont {M.~A.}\ \bibnamefont {{Dobbs}}}, \bibinfo {author}
  {\bibfnamefont {W.}~\bibnamefont {{Everett}}}, \bibinfo {author}
  {\bibfnamefont {J.}~\bibnamefont {{Gallicchio}}}, \bibinfo {author}
  {\bibfnamefont {E.~M.}\ \bibnamefont {{George}}}, \bibinfo {author}
  {\bibfnamefont {A.}~\bibnamefont {{Gilbert}}}, \bibinfo {author}
  {\bibfnamefont {N.}~\bibnamefont {{Gupta}}}, \bibinfo {author} {\bibfnamefont
  {N.~W.}\ \bibnamefont {{Halverson}}}, \bibinfo {author} {\bibfnamefont
  {N.}~\bibnamefont {{Harrington}}}, \bibinfo {author} {\bibfnamefont {G.~C.}\
  \bibnamefont {{Hilton}}}, \bibinfo {author} {\bibfnamefont {G.~P.}\
  \bibnamefont {{Holder}}}, \bibinfo {author} {\bibfnamefont {W.~L.}\
  \bibnamefont {{Holzapfel}}}, \bibinfo {author} {\bibfnamefont {J.~D.}\
  \bibnamefont {{Hrubes}}}, \bibinfo {author} {\bibfnamefont {N.}~\bibnamefont
  {{Huang}}}, \bibinfo {author} {\bibfnamefont {J.}~\bibnamefont {{Hubmayr}}},
  \bibinfo {author} {\bibfnamefont {K.~D.}\ \bibnamefont {{Irwin}}}, \bibinfo
  {author} {\bibfnamefont {L.}~\bibnamefont {{Knox}}}, \bibinfo {author}
  {\bibfnamefont {A.~T.}\ \bibnamefont {{Lee}}}, \bibinfo {author}
  {\bibfnamefont {D.}~\bibnamefont {{Li}}}, \bibinfo {author} {\bibfnamefont
  {A.}~\bibnamefont {{Lowitz}}}, \bibinfo {author} {\bibfnamefont {J.~J.}\
  \bibnamefont {{McMahon}}}, \bibinfo {author} {\bibfnamefont {S.~S.}\
  \bibnamefont {{Meyer}}}, \bibinfo {author} {\bibfnamefont {L.~M.}\
  \bibnamefont {{Mocanu}}}, \bibinfo {author} {\bibfnamefont {J.}~\bibnamefont
  {{Montgomery}}}, \bibinfo {author} {\bibfnamefont {A.}~\bibnamefont
  {{Nadolski}}}, \bibinfo {author} {\bibfnamefont {T.}~\bibnamefont
  {{Natoli}}}, \bibinfo {author} {\bibfnamefont {J.~P.}\ \bibnamefont
  {{Nibarger}}}, \bibinfo {author} {\bibfnamefont {G.}~\bibnamefont {{Noble}}},
  \bibinfo {author} {\bibfnamefont {V.}~\bibnamefont {{Novosad}}}, \bibinfo
  {author} {\bibfnamefont {S.}~\bibnamefont {{Padin}}}, \bibinfo {author}
  {\bibfnamefont {S.}~\bibnamefont {{Patil}}}, \bibinfo {author} {\bibfnamefont
  {C.}~\bibnamefont {{Pryke}}}, \bibinfo {author} {\bibfnamefont {J.~E.}\
  \bibnamefont {{Ruhl}}}, \bibinfo {author} {\bibfnamefont {B.~R.}\
  \bibnamefont {{Saliwanchik}}}, \bibinfo {author} {\bibfnamefont {K.~K.}\
  \bibnamefont {{Schaffer}}}, \bibinfo {author} {\bibfnamefont
  {C.}~\bibnamefont {{Sievers}}}, \bibinfo {author} {\bibfnamefont
  {G.}~\bibnamefont {{Smecher}}}, \bibinfo {author} {\bibfnamefont {A.~A.}\
  \bibnamefont {{Stark}}}, \bibinfo {author} {\bibfnamefont {C.}~\bibnamefont
  {{Tucker}}}, \bibinfo {author} {\bibfnamefont {K.}~\bibnamefont
  {{Vanderlinde}}}, \bibinfo {author} {\bibfnamefont {T.}~\bibnamefont
  {{Veach}}}, \bibinfo {author} {\bibfnamefont {J.~D.}\ \bibnamefont
  {{Vieira}}}, \bibinfo {author} {\bibfnamefont {G.}~\bibnamefont {{Wang}}},
  \bibinfo {author} {\bibfnamefont {N.}~\bibnamefont {{Whitehorn}}}, \bibinfo
  {author} {\bibfnamefont {W.~L.~K.}\ \bibnamefont {{Wu}}}, \bibinfo {author}
  {\bibfnamefont {V.}~\bibnamefont {{Yefremenko}}},\ and\ \bibinfo {author}
  {\bibnamefont {{SPTpol Collaboration}}},\ }\bibfield  {title} {\bibinfo
  {title} {{Measurements of B -mode polarization of the cosmic microwave
  background from 500 square degrees of SPTpol data}},\ }\href
  {https://doi.org/10.1103/PhysRevD.101.122003} {\bibfield  {journal} {\bibinfo
   {journal} {\prd}\ }\textbf {\bibinfo {volume} {101}},\ \bibinfo {eid}
  {122003} (\bibinfo {year} {2020})},\ \Eprint
  {https://arxiv.org/abs/1910.05748} {arXiv:1910.05748 [astro-ph.CO]}
  \BibitemShut {NoStop}%
\bibitem [{\citenamefont {{The Simons Observatory
  collaboration}}(2019)}]{SimonsObservatory}%
  \BibitemOpen
  \bibfield  {author} {\bibinfo {author} {\bibnamefont {{The Simons Observatory
  collaboration}}},\ }\bibfield  {title} {\bibinfo {title} {{The Simons
  Observatory}},\ }in\ \href@noop {} {\emph {\bibinfo {booktitle} {BAAS}}},\
  Vol.~\bibinfo {volume} {51}\ (\bibinfo {year} {2019})\ p.\ \bibinfo {pages}
  {147},\ \Eprint {https://arxiv.org/abs/1907.08284} {arXiv:1907.08284
  [astro-ph.IM]} \BibitemShut {NoStop}%
\bibitem [{\citenamefont {{CMB-S4 Collaboration}}(2019)}]{CMBS4}%
  \BibitemOpen
  \bibfield  {author} {\bibinfo {author} {\bibnamefont {{CMB-S4
  Collaboration}}},\ }\bibfield  {title} {\bibinfo {title} {{CMB-S4 Science
  Case, Reference Design, and Project Plan}},\ }\href@noop {} {\bibfield
  {journal} {\bibinfo  {journal} {arXiv e-prints}\ ,\ \bibinfo {eid}
  {arXiv:1907.04473}} (\bibinfo {year} {2019})},\ \Eprint
  {https://arxiv.org/abs/1907.04473} {arXiv:1907.04473 [astro-ph.IM]}
  \BibitemShut {NoStop}%
\bibitem [{\citenamefont {{LiteBIRD Collaboration}}(2022)}]{PtepLB}%
  \BibitemOpen
  \bibfield  {author} {\bibinfo {author} {\bibnamefont {{LiteBIRD
  Collaboration}}},\ }\bibfield  {title} {\bibinfo {title} {{Probing Cosmic
  Inflation with the LiteBIRD Cosmic Microwave Background Polarization
  Survey}},\ }in\ \href {https://doi.org/10.1117/12.2563050} {\emph {\bibinfo
  {booktitle} {PTEP}}},\ \bibinfo {series} {PTEP}, Vol.\ \bibinfo {volume}
  {11443}\ (\bibinfo {year} {2022})\ p.\ \bibinfo {pages} {114432F},\ \Eprint
  {https://arxiv.org/abs/2101.12449} {arXiv:2101.12449 [astro-ph.IM]}
  \BibitemShut {NoStop}%
\end{thebibliography}%

\appendix
\section{MCMC chain plots and tables\label{sec:appendix}}
We display here information about the chains and their convergence derived using {\sc montepython} in subsection \ref{ssec:conv}. Full plots using {\sc getdist} and including the cosmological parameters are also displayed in subsection \ref{ssec:fullcorner}. 
\subsection{Convergence information}\label{ssec:conv}
Tables \ref{tab:bsbm-tab-R-1}, \ref{tab:zm-tab-R-1}, and \ref{tab:OP-full} list the convergence of the chain specified through the Gelman-Rubin criterion $|R-1|$. Values much smaller than $0.1$ typically indicate well converged chains, which is the case for all parameters across all chains. We further show for reproducibility the initial guesses for mean and standard deviation ($\mu_0$ and $\sigma_0$\,, respectively) to reduce the burn in of the MCMC chains. We stress that these are not Gaussian priors imposed on our parameters.

\begin{table}[h]
    \centering
    \caption{Complementary information for the free parameters of the BSBM model (not including nuisance parameters). In the first column we show the parameter name, in the second column the $|R-1|$ Gelman-Rubin convergence criterion, and in the third and fourth column the mean and standard deviation to initialize the chains with (these are not priors).  Total number of accepted steps: 3736795 for 16 chains.}
    \begin{tabular}{l|c|c|c}
\toprule
            \textbf{Parameter}  & 
            \textbf{R-1} & {\boldmath$\mu_0$} & {\boldmath $\sigma_0$}\\
\toprule
                {\boldmath$100 \omega_b$} & $ 0.005936$ & 2.2377 &   0.015 \\
\midrule
                {\boldmath$\omega_{\rm cdm}$} & 0.003368 & 0.12010 & 0.0013\\
\midrule
                {\boldmath$H_0$} & $0.002793$ & 67.8 &0.5 \\
\midrule
                {\boldmath$ln10^{10}A_{s }$} & $0.005616$ &3.0447 &0.015 \\
\midrule
                    {\boldmath$n_{s }  $} & $0.005490$&0.9659&0.0042 \\
\midrule
                  {\boldmath$z_{reio } $} & $ 0.006086 $& 8&0.5 \\
\midrule
                  {\boldmath$\zeta $ (ppm)} & $0.002793 $ & 0& 0.1 \\
\bottomrule
\end{tabular}
\label{tab:bsbm-tab-R-1}
\\ 
$-\ln{\cal L}_\mathrm{min} =2047.22$, 
\end{table}
\begin{table}[h]
    \centering
    \caption{Same as table \ref{tab:bsbm-tab-R-1}, but for the O$\&$P model universally coupled to matter. Total number of accepted steps: 1174065 for  14 chains.}
    \begin{tabular}{l|c|c|c}

\toprule
            \textbf{Parameter}  & 
            \textbf{R-1}& {\boldmath$\mu_0$} & {\boldmath $\sigma_0$}\\
\toprule
                {\boldmath$100 \omega_b$} & 0.013401 & 2.2377 &   0.015\\
\midrule
                {\boldmath$\omega_{\rm cdm}$} &  0.018249 & 0.12010 & 0.0013 \\
\midrule
                {\boldmath$H_0$} & 0.020532 & 67.8 &0.5 \\
\midrule
                {\boldmath$ln10^{10}A_{s }$} & 0.013956 &3.0447 &0.015 \\
\midrule
                    {\boldmath$n_{s }  $} & $0.033703 $&0.9659&0.0042 \\
\midrule
                  {\boldmath$z_{reio } $} &$ 0.010774 $& 8&0.5 \\
\midrule
                  {\boldmath$\eta_m $} (ppm) & $0.020532$& 0& $10^{-6}$\\
\midrule
                  {\boldmath$\eta_\Lambda $} (ppm) & $0.017963 $&0& 0.01  \\
\bottomrule
\end{tabular}
\label{tab:zm-tab-R-1}
\\ 
$-\ln{\cal L}_\mathrm{min} =2048.09$ 
\end{table}
\begin{table}[h!]
    \centering
    \caption{Same as table \ref{tab:bsbm-tab-R-1}, but for the full O$\&$P model. Total number of accepted steps: 1143660  for 28 chains.}
    \begin{tabular}{l|c|c|c}
\toprule
            \textbf{Parameter}  & 
            \textbf{R-1}& {\boldmath$\mu_0$} & {\boldmath $\sigma_0$}\\
\toprule
                {\boldmath$100 \omega_b$} & 0.007294 & 2.2377 &   0.015 \\
\midrule
                {\boldmath$\omega_{\rm cdm}$} & $ 0.005807 $& 0.12010 & 0.0013 \\
\midrule
                {\boldmath$H_0$} & $0.005984$& 67.8 &0.5 \\
\midrule
                {\boldmath$ln10^{10}A_{s }$} & 0.006341 &3.0447 &0.015 \\
\midrule
                    {\boldmath$n_{s }  $} & $ 0.005657 $&0.9659&0.0042 \\
\midrule
                  {\boldmath$z_{reio } $} & 0.005787 & 8&0.5 \\
\midrule
                  {\boldmath$\eta_\chi $} (ppm) & 0.005984 &0& 0.01 \\
\midrule
                  {\boldmath$\eta_b $} (ppm) & 0.005984  & 0& $10^{-6}$ \\
\midrule
                  {\boldmath$\eta_\Lambda $} (ppm) & 0.037416&0& 0.01 \\
\bottomrule
\end{tabular}
\label{tab:OP-full}
\\ 
$-\ln{\cal L}_\mathrm{min} =2047.41$ 
\end{table}

\pagebreak
\subsection{Full corner plots}\label{ssec:fullcorner}

\begin{figure*}[t]
    \includegraphics[scale=0.42]{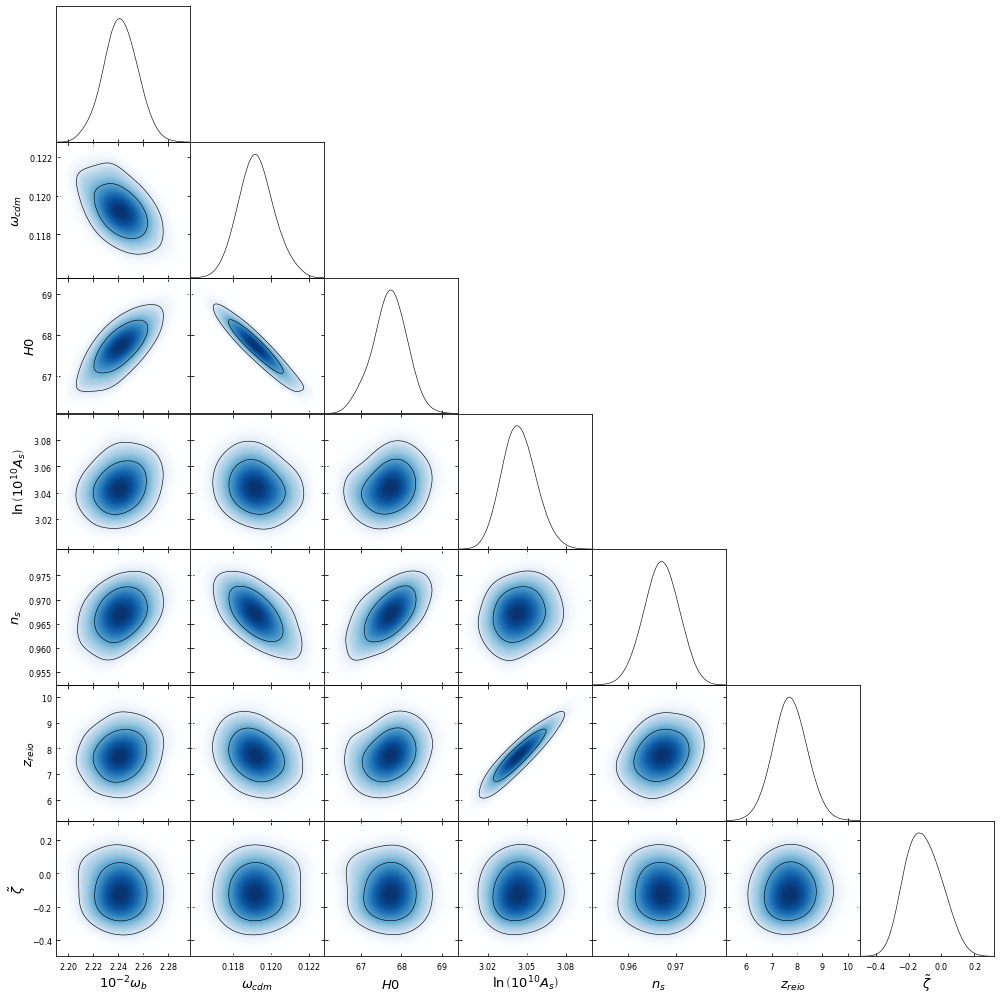}
    \caption{Contour plots for the full parameter space (Bekenstein + cosmology) in the case of the BSBM model. Note that $\tilde{\zeta}$ is presented in ppm.}
    \label{fig:bsbm-full}
\end{figure*}
In this section we display the full corner plots for the three models analyzed in section \ref{sec:results}. The very good convergence is immediately apparent in the figures, as well as the lack of any significant degeneracy with the parameters of the given model. A very attentive reader might notice that the correlations of the Bekenstein and cosmological parameters are not always prefect ellipses, hence indicating the non-triviality of such a study.
\begin{figure*}[h]
    \includegraphics[scale=0.42]{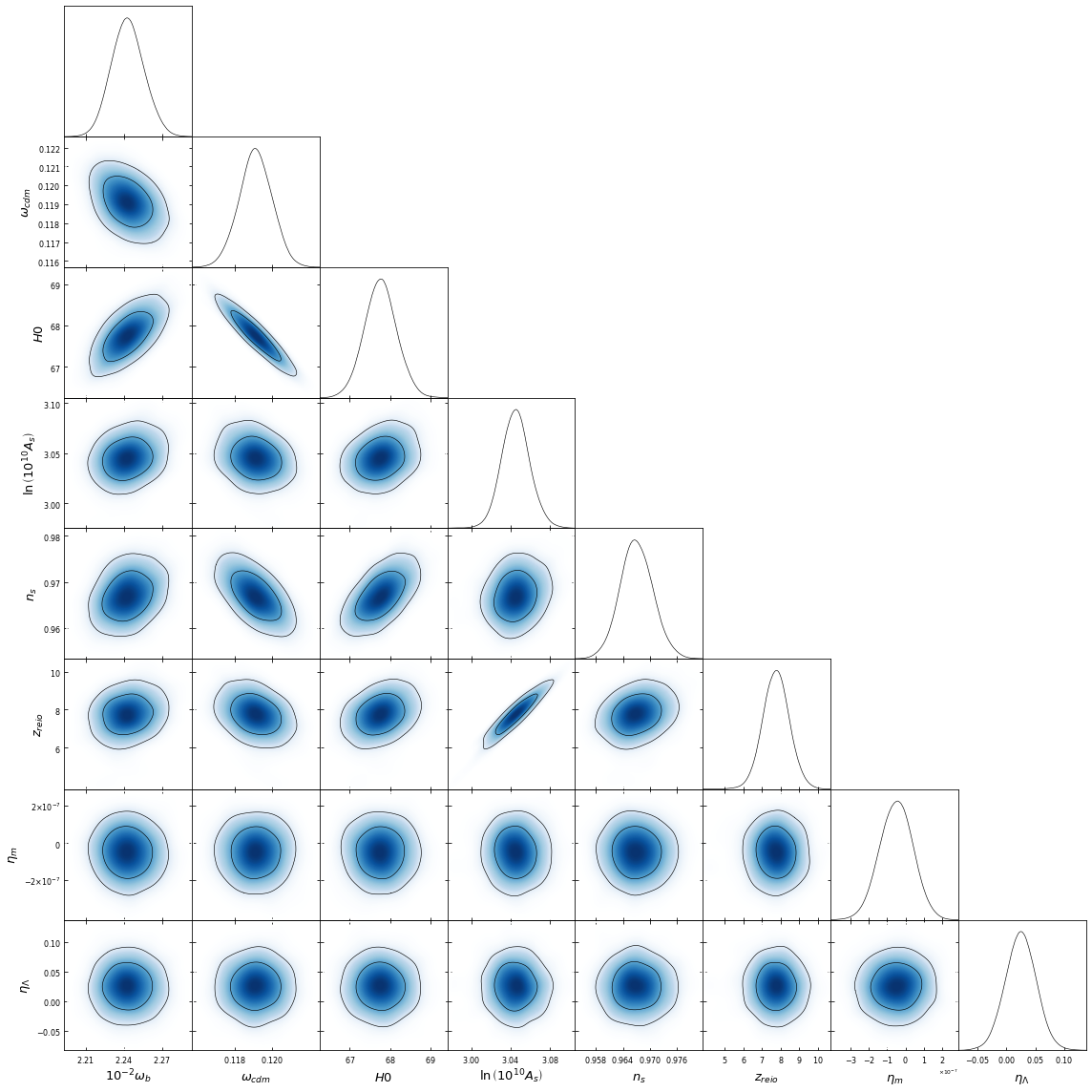}
    \caption{Contour plots for the full parameter space (Bekenstein + cosmology) in the case of the O$\&$P model universally coupled to gravity. Note that $\eta_m$ and $\eta_{\Lambda}$ are presented in ppm (and for $\eta_m$ there is an additional scaling of $10^{-6}$).}
    \label{fig:zm-full}
\end{figure*}
\begin{figure*}[h]
    \includegraphics[scale=0.4]{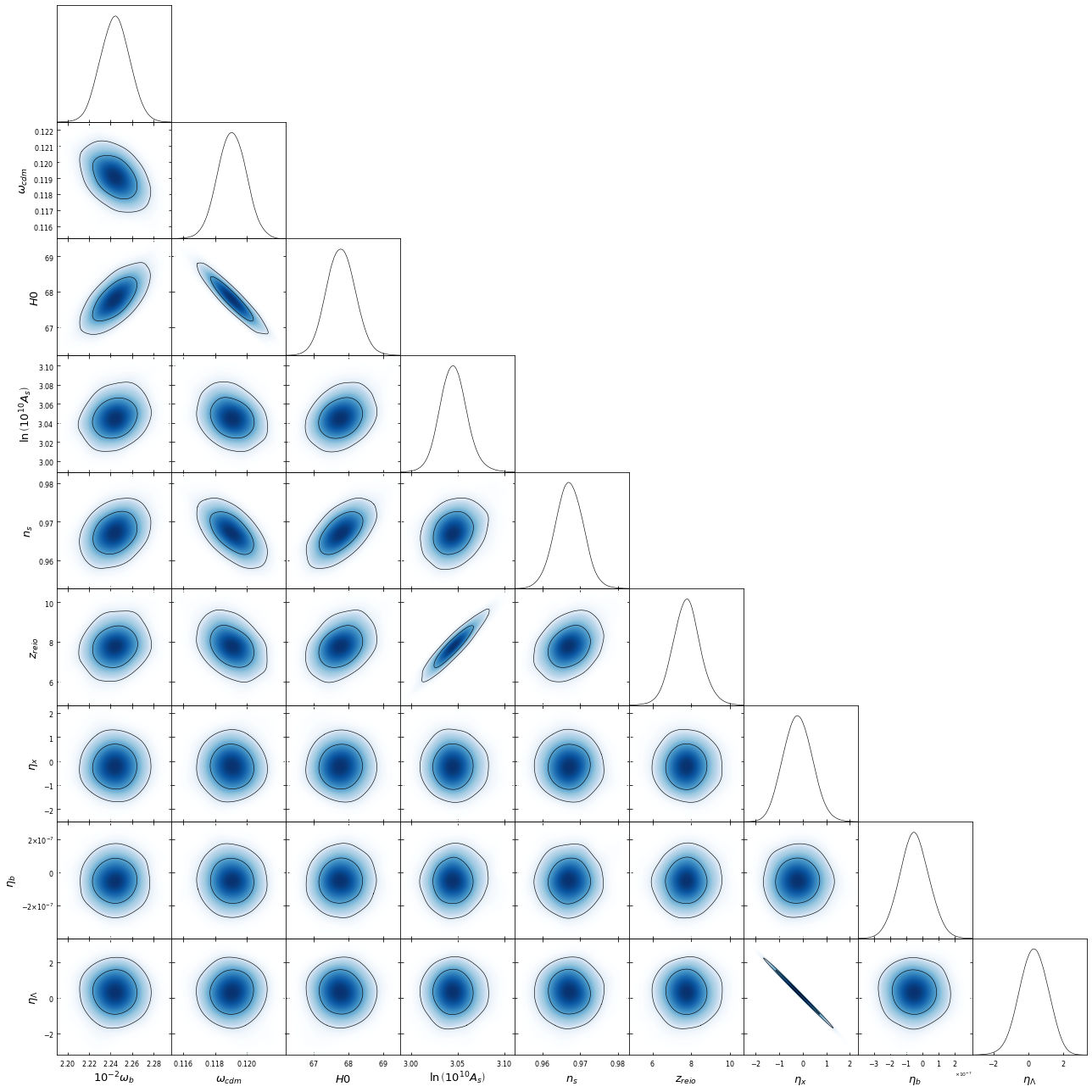}
    \caption{Contour plots for the full parameter space (Bekenstein + cosmology) in the case of the full O$\&$P model. Note that $\eta_{\chi}$, $\eta_b$ and $\eta_{\Lambda}$ are presented in ppm (and for $\eta_b$ there is an additional scaling of $10^{-6}$).}
    \label{fig:bkst-full}
\end{figure*}

\end{document}